\documentclass[
aps,
prd,
preprint,
showpacs,
eqsecnum,
nofootinbib,
]{revtex4}

\usepackage{amsbsy}
\usepackage{amsfonts}
\usepackage{amssymb}
\usepackage{amsmath}
\usepackage{graphicx}
\usepackage{subfigure}

\DeclareMathOperator{\li}{Li}
\newcommand{\mathsym}[1]{{}}

\def\IP{\mathbb{P}}

\def\IO{\mathbb{O}}
\def\IE{\mathbb{E}}

\def\boldA{\boldsymbol{A}}

\def\id{\protect{{1 \kern-.28em {\rm l}}}}

\def\be{\begin{eqnarray}}
\def\ee{\end{eqnarray}}

\def\Tr{{\rm Tr}}
\def\MHVbar{${\overline{\rm MHV}}$}
\def\nn{\nonumber}
\def\NeqFour{${\cal N}=4$}
\def\Split{{\rm Split}}
\def\spa#1.#2{\left\langle#1\,#2\right\rangle}
\def\spb#1.#2{\left[#1\,#2\right]}
\def\sandmx#1.#2.#3{%
\left\langle#1{\vphantom1}\right|{#2}\left|#3\right]}%
\def\delt#1{\delta^{(#1)}}
\def\eps{\epsilon}
\def\Ord{{\cal O}}
\def\tlambda{\tilde\lambda}

\def\tp{\!+\!}
\def\sA{{\cal A}}
\def\Int#1{I^{(#1)}}

\def\hs{\hskip .4 cm}

\begin{document}

\preprint{Saclay/IPhT--T10/096}
\preprint{Brown-HET-1607}

\title{The Six-Point NMHV amplitude in Maximally Supersymmetric Yang-Mills Theory}

\author{D.~A.~Kosower}
\affiliation{Institut de Physique Th\'eorique, CEA--Saclay, F--91191
  Gif-sur-Yvette cedex, France}
\email{David.Kosower@cea.fr}
\author{R.~Roiban}
\affiliation{Department of Physics, Pennsylvania State University,
  University Park, PA 16802, USA}
\email{radu@phys.psu.edu}
\author{C.~Vergu}
\affiliation{Department of Physics, Brown University, Box 1843, Providence, RI 02912, USA}
\email{Cristian_Vergu@brown.edu}

\begin{abstract}
 
 We present an integral representation for the parity-even part of the
 two-loop six-point planar NMHV amplitude in terms of Feynman integrals which
 have simple transformation properties under the dual conformal symmetry. 
 We probe the dual conformal properties of the amplitude 
 numerically, subtracting the known infrared
 divergences. We find that the subtracted amplitude
 is invariant under dual conformal
 transformations, confirming existing conjectures through
 two-loop order. We also discuss the all-loop structure of the
 six-point NMHV amplitude and give a parametrization whose dual
 conformal invariant building blocks have a simple
 physical interpretation.
\end{abstract}


\maketitle

\newpage

\section{Introduction}

What is the scattering matrix of a matter-coupled gauge theory? In
general, this is a hard question involving both conceptual and
technical subtleties.  Nevertheless, scattering amplitudes enjoy a
much simpler structure than implied by their expansion in terms of
Feynman diagrams.  For some theories, additional off-shell and
on-shell symmetries simplify the amplitudes enormously.  The further
simplification exhibited in the planar (or infinite-color) limit may
even allow a complete answer to the question.

The \NeqFour{} or maximally supersymmetric Yang--Mills theory (MSYM)
may be such a theory.  The simplifications inherent in the larger
symmetry have already allowed explicit calculations of scattering
amplitudes well beyond those for other theories.  At weak coupling,
advances in multi-loop and multi-leg
calculations~\cite{NeqFourOneLoop,Bern:1994cg,UnitarityII,BCFUnitarity,
  Drummond:2008bq,Bern:2009xq,OnShell} have opened the possibility of
probing the structure of the scattering matrix to high order in
perturbation theory.  The BDS conjecture~\cite{BDS} for the all-loop
resummation of maximally helicity-violating (MHV) amplitudes (based on
an earlier relation~\cite{ABDK} linking one- and two-loop amplitudes)
provides an example of possible structures that can emerge.  At strong
coupling, the leading expansion of scattering amplitudes has been
computed using the AdS/CFT correspondence~\cite{AdS/CFT} by Alday,
Gaiotto, Maldacena, Sever, and Vieira~\cite{AM1,AM2,AGM,AMSV}.  This
two-sided approach, together with the recent developments in the
evaluation of scattering amplitudes at strong coupling for any number
of external legs and the realization that the relation between certain
scattering amplitudes and null polygonal Wilson loops carries over
from strong coupling~\cite{Tduality} to the weak-coupling
regime~\cite{DKS,BHT,DHKS2loops,Bern:2008ap,WilsonCorrelationAmplitudeLink}
offer circumstantial evidence that the \NeqFour{} super-Yang-Mills
theory may ultimately be solvable in the planar limit.

Gluon MHV amplitudes, with two external legs of negative helicity and
the rest of positive helicity,
are the simplest amplitudes in a gauge theory.  They are particularly
simple in the planar limit of MSYM, where they are determined by a single
helicity structure, accompanied by a
function of scalar and pseudo-scalar momentum invariants, to all
orders in perturbation theory.  This simplicity is of course shared
by their parity conjugates, the \MHVbar{} amplitudes.
The structure of the remaining non-MHV amplitudes is
more complicated.   At one loop, explicit expressions are 
known~\cite{NMHV7point,OneLoopNMHV}
at arbitrary multiplicity
for next-to-MHV (NMHV) gluon amplitudes, with three gluons of negative
helicity from the rest.  These are already more intricate,
with the number of independent helicity structures growing cubicly with
the number of external legs, each multiplied by an independent function
of scalar and pseudo-scalar invariants.  How does this structure
generalize to higher loops?  No explicit expressions are known to date
for higher-loop non-MHV amplitudes.  In this paper, we take a first
step towards filling this gap, computing the parity-even part of the two-loop
six-gluon NMHV amplitude.  These results were first reported 
in ref.~\cite{QueenMaryTalk}. This amplitude, which comes with three
inequivalent helicity configurations, is the simplest non-MHV amplitude.

General results on the structure of infrared divergences in massless
gauge theories suggest on one hand, that the divergent terms have
a simple iterative structure; and on the other, that all planar amplitudes
with fixed number of external legs share the same structure
of infrared-divergent
terms.  Together with the structure of the splitting amplitudes, this
implies that a natural way to extract infrared-finite quantities from
non-MHV amplitudes is to divide them by the MHV amplitudes with the
same number of external legs. 
Drummond, Henn, Korchemsky,
and Sokatchev (DHKS) showed
to one-loop order that this ratio is not only 
finite but also dual conformal invariant for 
NMHV amplitudes~\cite{Drummond:2008vq,Drummond:2008bq}
and conjectured that the same holds to all orders 
for all non-MHV amplitudes~\cite{Drummond:2008vq}.
%
\iffalse
Drummond, Henn, Korchemsky,
and Sokatchev (DHKS) showed
at one-loop order, and conjectured to all orders that
 the this ratio is not only finite but also dual
conformal invariant for NMHV
amplitudes~\cite{Drummond:2008vq}. 
\fi
%
Here we clarify and test this conjecture
to two-loop order for the six-point amplitude.
This test requires the use of Dixon and Schabinger's
result~\cite{SD} for the $\Ord(\eps)$ terms in the one-loop NMHV six-point
amplitude.

As an intriguing consequence of the semiclassical approach of 
Alday and Maldacena~\cite{AM1},
anticipated by the structure of flat space string theory scattering
amplitudes at high energy and fixed 
angles~\cite{GrossMende}, to leading order
in the strong coupling expansion all scattering amplitudes are (in
a certain sense) insensitive to the flavor and polarization of external
legs.  While quantum corrections are likely to alter this conclusion,
this structure is surprising from the standpoint of the intricate
analytic structure of the weak-coupling scattering matrix.

The arguments of Alday and Maldacena led to the identification
\cite{DKS, BHT} of a surprising relation between one-loop MHV
amplitudes and the one-loop expectation value of special null
polygonal Wilson loops. This relation was shown to hold at two loops
as well for four-, five-~\cite{DHKS2loops} and six-particle scattering
amplitudes~\cite{Bern:2008ap,DHKS6pt}. Integral representations of
higher-point two-loop MHV amplitudes are also known~\cite{CVn};
comparison with Wilson-loop expectation values~\cite{BabisetAl} is
hindered, however, by the complexity of evaluating the required 
higher-point two-loop
Feynman integrals.  Dual conformal 
symmetry~\cite{MagicIdentities,Drummond:2008vq,BHT,DHKS2loops,CWI}
plays an important role in the 
relation between MHV scattering amplitudes and Wilson loops.
%\cite{DKS, DHKS2loops} 
This symmetry is manifest for the integrands
of both MHV scattering amplitudes and Wilson loops, but it is broken
by the dimensional regulator. Dual-conformal invariants can be
constructed by using the general structure of divergent terms.
A particular pattern of spontaneous breaking of the gauge group
provides an alternative regularization in which this symmetry is restored 
through natural transformations of the regulator~\cite{AHPS}.
%
\iffalse
preserves this  symmetry~\cite{AHPS}.
\fi
%
%%
%This symmetry is manifest for the relevant Wilson loops. The available MHV 
%amplitudes have been shown to exhibit it as well. 
%%

It would be interesting to understand whether non-MHV amplitudes also
exhibit a similar presentation in terms of Wilson loops. A necessary
condition is that they exhibit dual conformal invariance upon
extraction of infrared divergences.
It is possible to argue that, to all orders in the loop expansion,
four-dimensional cuts of any planar scattering amplitude in \NeqFour~SYM, in particular non-MHV amplitudes, have this symmetry. Hints in
this direction also come from the Grassmannian interpretation of leading
singularities; in that framework it was shown~\cite{Bullimore:2009cb,
  Korchemsky:2010ut} that leading singularities are dual conformally
invariant.
Whether this symmetry survives in the complete amplitude, in the
presence of the terms not constructible from
four-dimensional cuts, is an open question.  Here we will see that
the parity-even part of the six-point NMHV amplitudes can be expressed in
terms of pseudo-conformal integrals, {\it i.e.\/} dimensionally regulated
integrals that are invariant under dual
conformal transformations
when continued off-shell.

While the structure of collinear limits of non-MHV amplitudes is
somewhat more intricate than those of MHV amplitudes,
the former are governed by the same splitting
amplitudes as the latter.  The iteration
relation for MHV amplitudes~\cite{ABDK} suggests 
that one can capture both the infrared-divergent parts
 of non-MHV amplitudes, as well as the amplitudes' behavior under collinear
limits, via an exponentiation ansatz for all the scalar functions that
characterize them.  This is similar in spirit to the 
BDS~\cite{BDS} exponentiation ansatz for MHV amplitudes.  
Such an ansatz is not expected
to hold to all orders.  Departures from it are characterized by 
dual conformal invariant functions which have properties
analogous to the MHV remainder function~\cite{Bern:2008ap, DHKS6pt}.

We perform the calculation using the generalized unitarity-based method,
employing a variety of four-dimensional and $D$-dimensional cuts to
express the amplitude in terms of six-point two-loop Feynman
integrals.  The four-dimensional cuts are evaluated in on-shell
superspace~\cite{Nair:1988bq}. This approach automatically takes into
account supersymmetry relations between different components of cuts
and also offers guidance in organizing the calculation. We find that
the (appropriately defined) parity-even part of the six-point
amplitude may be expressed as a sum of pseudo-conformal
integrals~\cite{MagicIdentities}, in close analogy with the four-point
amplitude through five loops~\cite{GSB,BRY,BDS,BCDKS,FiveLoop} and the
parity-even part of the five-point amplitude through two
loops~\cite{FiveGluonOneLoop,BDDKSelfDual,TwoLoopFiveA,Bern:2006vw}.
There are some additional integrals in the one- and two-loop six-point
amplitudes, whose pseudo-conformal nature is less clear.  Their
integrands vanish as $D\to4$, yet their integrals can be nonvanishing
in this limit.
We evaluate the integrals using the {\tt AMBRE}~\cite{AMBRE} and
{\tt MB}~\cite{MB} packages and compute the amplitude numerically at several
kinematic points, related in pairs by dual conformal transformations.  
The infrared singularities of our expression have the structure
expected from general
considerations~\cite{KnownIR,KorchemskyMarchesini}. 
We have tested numerically the dual-conformal
properties of the various finite functions that can be constructed
from the six-point NMHV amplitude.

The paper is organized as follows. 
We review the tree-level and one-loop six-point amplitudes in
section~\ref{review}, along with their superspace presentation and
their conjectured properties. Most importantly, we identify a
canonical separation of the six-point NMHV amplitude into parity-even
and parity-odd components.  We expect this separation to extend to all
orders in perturbation theory.
In section~\ref{higher_loop_structure}, we discuss 
 the expected structure of the six-point NMHV amplitude
to all loop orders, based on our calculation using generalized unitarity. 
We introduce certain finite functions that characterize the amplitude and 
are expected to be invariant under dual conformal transformations. 
In section~\ref{TheCalculation}, we describe some of the details of
our calculation. We use
a superspace version of the generalized unitarity
method.  We discuss some of the subtle points, and give
details on the calculation of two important
cuts.
In section~\ref{2loopintegrands}, we present an integral
representation of the even part of the two-loop six-point NMHV
amplitude. For completeness we also list the even part of the two-loop
six-point MHV amplitude in our notation.
We proceed in section~\ref{results} to analyze our analytic and
numerical results for the amplitude, and to test the dual
conformal-symmetry
properties of the various functions that have been conjectured to be
invariant under dual conformal transformations.
We give our conclusions and 
a selection of open problems in section~\ref{summary}.

\noindent
{\bf Note added:} As the writing of this paper was being completed we
received ref.~\cite{AllLoopLS} in which an alternative presentation of
the six-point NMHV amplitude was proposed as a consequence of a
generalization of the Grassmannian duality for leading singularities
to the full amplitude. The result also contains a proposal for
the parity-odd part of the amplitude.  Unlike our result,
it is expressed in terms of a basis
of chiral, tensor integrals written in momentum-twistor space.

%%%%%%%%%%%%%%%%%%%%%%%%%%%%%%%%%%%%%%%%%%%%%
\section{Review \label{review}  }

The $n$-point $L$-loop planar (leading-color) contributions
to scattering amplitudes of an $SU(N_c)$
gauge theory with fields in the adjoint representation 
% ??? and couplings depending on the antisymmetric structure constants 
may be written
as\footnote{We normalize the classical action so that the only
coupling constant dependence is an overall factor of $g_{YM}^{-2}$.}
\be
\boldA_n^{(L)}=a^L\sum_{\rho\in S_n/\mathbb{Z}_n} \Tr[T^{a_{\rho(1)}}\dots T^{a_\rho(n)}]
A_n^{(L)}(k_{\rho(1)},\varepsilon_{\rho(1)};\dots;k_{\rho(n)}, \varepsilon_{\rho(n)})\,,
\ee
where we follow the normalization conventions\footnote{This definition of
the loop expansion parameter extracts the complete dependence on the Euler
constant from the momentum integrals.} of ref.~\cite{Bern:2008ap}
(which differ from those used in
refs.~\cite{Bern:1994cg, Drummond:2008vq}).
The loop expansion parameter 
$a$ is,
\begin{equation}
  a=(4\pi e^{-\gamma})^{-\eps}\frac{\lambda}{8\pi^2}
=(4\pi e^{-\gamma})^{-\eps}\frac{g_{\rm YM}^2 N_c}{8\pi^2}
\,.
\label{eq:loop-exp-param}
\end{equation} 
Here $\lambda$ is the 't~Hooft coupling constant and
$\gamma$ is the Euler constant, $\gamma = -\Gamma'(1)$.  The sum runs
over all the noncyclic permutations of the external legs, each of
which carries momentum $k_i$ and a polarization vector $\varepsilon_i$.

Choosing a specific helicity and flavor configuration for the external
legs reduces
$A_n^{(L)}(k_{\rho(1)},\varepsilon_{\rho(1)};\dots;k_{\rho(n)},
\varepsilon_{\rho(n)})$ to a color-ordered partial amplitude.  Every
partial amplitude can be decomposed into a sum of terms, each of
which is a product of a function ensuring the correct transformation
properties of the amplitude under Lorentz transformations (henceforth
called ``spin factor'') and a (pseudo-)scalar function which may be
written as a sum of $L$-loop Feynman integrals (the ``loop factor'').
The spin factor is a rational function of the momentum spinors
$\lambda_i$ and $\tilde\lambda_i$ associated to the external legs; the
parity-even parts of the loop factor are functions of external Lorentz
invariants alone, while the parity-odd parts also depend on
Levi-Civita contractions of the external momenta.  One could of
course choose to re-express the Levi-Civita contractions in terms of
spinor variables.

MHV amplitudes have two negative-helicity, and any number of
positive-helicity, external legs.  These amplitudes in MSYM have the
simplest structure of all amplitudes: they have a single spin factor,
which is equal to the tree-level scattering amplitude.  Computing the
$L$-loop MHV amplitude thus amounts to finding the ratio
\be
M_n^{(L)} \equiv \frac{A_n^{(L),\text{MHV}}}{A_n^{(0),\text{MHV}}}\,.
\ee
CPT implies similar properties for the $\overline{\text{MHV}}$
amplitudes; they also contain a single spin factor which is the
tree-level amplitude and their scalar and pseudo-scalar factors are
obtained from corresponding MHV amplitude by a parity transformation.
(For alternative presentations of the parity-odd terms in $M_n$ in
terms of spinor variables, see refs.~\cite{Cachazo:2008vp,Spradlin:2008uu}.)

All-gluon NMHV amplitudes have three external legs of negative helicity,
and any number of positive helicity.  They are the next-simplest
amplitudes after the MHV ones.  The five-point NMHV amplitudes are 
$\overline{\text{MHV}}$; the simplest distinct ones 
appear for six external legs.  These have three independent helicity
configurations.  In contrast to the MHV amplitudes, NMHV amplitudes
contain several distinct spin factors; their forms
depend on the helicity configuration of the external legs.
As a consequence of relations between spin factors, there are many
possible presentations of the tree-level amplitudes. 
We can single
out a canonical form by constructing the corresponding one-loop
amplitude and taking the form that appears as the coefficient of
the double pole in the dimensional-regularization parameter
$\eps$. This relation~\cite{Giele:1991vf, Kunszt:1992tn,
  Giele:1993dj, Kunszt:1994mc} is guaranteed by the general theorems
governing the factorization of soft and collinear divergences. 
We will focus here on the
the six-point amplitude.

%%%%%%%%%%%%%%%%%%%%%%%%%%%%%%%%%

\subsection{The Six-Point Gluon Scattering Amplitude at One Loop}

All six-gluon NMHV amplitudes may be obtained by applying cyclic 
permutations and reflections to the three independent helicity
configurations $({+}{+}{+}{-}{-}{-})$, $({+}{+}{-}{+}{-}{-})$ and
$({+}{-}{+}{-}{+}{-})$. The one-loop amplitudes for these 
configurations were first obtained in ref.~\cite{Bern:1994cg} 
through $\Ord(\eps^0)$ (see also ref.~\cite{OneLoopNMHV}).
They can be expressed in terms of
three different spin factors.  
The spin factors for the 
`split-helicity' configuration $({+}{+}{+}{-}{-}{-})$ are,
  \be
     \label{eq:b1}
    B_1 & = & i \frac{s_{123}^3}{\left\langle 1 2\right\rangle \left\langle 2
        3\right\rangle \langle 1(2+3)4] \langle 3(1+2)6] \left[4
        5\right] \left[5 6\right]}\,,\\
    \label{eq:b2}
    B_2 & = &
      i \frac{\langle 4(2+3)1]^3}{\left\langle 2 3\right\rangle
        \left\langle 3 4\right\rangle
        \langle 2(3+4)5] \left[5 6\right] \left[6 1\right]
        s_{234}}
        + i \frac{\left\langle 5 6\right\rangle ^3 \left[2
          3\right]^3}{\left\langle 6 1\right\rangle  \left\langle
          1(2+3)4\right] \left\langle 5(3+4)2\right] \left[3 4\right]
        s_{234}}\,, \\
    \label{eq:b3}
  B_3  &=& 
      i \frac{\langle 6(1+2)3]^3}{\left\langle 6 1\right\rangle
        \left\langle 1 2\right\rangle \langle 2(1+6)5] \left[3
          4\right] \left[4 5\right] s_{345}}
      + i \frac{\left\langle 4 5\right\rangle ^3 \left[1
          2\right]^3}{\left\langle 3 4\right\rangle  \left\langle
          3(1+2)6\right] \left\langle 5(1+6)2\right] \left[6 1\right]
        s_{345}}\,;
  \ee
we refer the reader to the original paper~\cite{Bern:1994cg} for the
spin factors of the other independent helicity
configurations. In all cases, the spin factors are uniquely determined
by cuts in three-particle invariants.

  \begin{figure}
    \centering
    \includegraphics{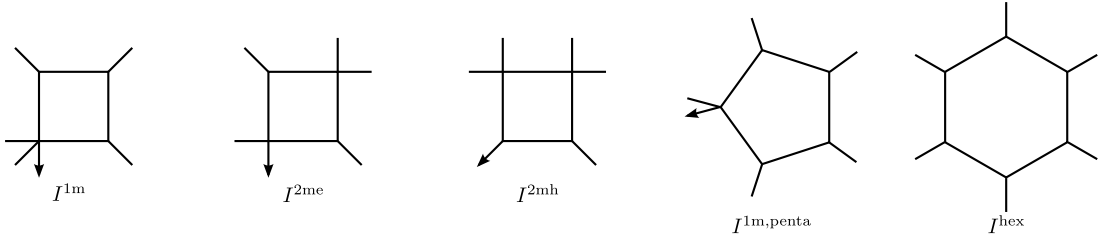}
    \caption{The integrals contributing to the six-point one-loop MHV
      and NMHV amplitudes.  An arrow marks the leg with momentum
      $k_1$; the remaining momenta follow clockwise.  The
      one-mass box $I^{\text{1m}}$ and two-mass easy $I^{\text{2me}}$
      integrals contribute to the MHV amplitude and the one-mass box
      $I^{\text{1m}}$ and two-mass hard $I^{\text{2mh}}$ integrals contribute to
      the NMHV amplitude.  The one-mass pentagon $I^{\text{1m,penta}}$
      and the hexagon $I^{\text{hex}}$ have numerator factors of
      $\mu^2$ (the square of the $(-2 \eps)$-dimensional
      components of the loop momentum), and hence are finite.  
      They contribute to both the
      MHV and NMHV amplitudes only at $\Ord(\eps)$ and higher
      ($I^{\text{hex}}$ contributes to the even parts while
      $I^{\text{1m,penta}}$ contributes to the odd parts).}
    \label{fig:1loop-integrals}
  \end{figure}

The six-point
one-loop NMHV amplitude for the $({+}{+}{+}{-}{-}{-})$ 
helicity configuration is given by,
\begin{equation}
  \label{eq:1loop-pppmmm}
  A_6^{(1),\text{NMHV}}({+}{+}{+}{-}{-}{-}) = \frac 1 2 \left(B_1
    W_1^{(1)} + B_2 W_2^{(1)} + B_3 W_3^{(1)}\right)+{\cal O}(\eps)\,,
\end{equation} 
where,
%We will explain these conventions later on.
\begin{equation}
  \label{eq:w1-1}
  W_1^{(1)} = - \frac 1 2 \sum_{\sigma \in \mathcal{S}_1} \left(\frac
    1 2 s_{45} s_{56} I^{\text{1m}}(\sigma) + \frac 1 2 s_{61} s_{123}
    I^{\text{2mh}}(\sigma)\right)+{\cal O}(\epsilon)\,;
\end{equation} 
and the sum runs over the permutations,
\begin{align}
  \label{eq:perm1}
  \mathcal{S}_1 &= \lbrace (123456), (321654), (456123),
  (654321)\rbrace\,.
\end{align}
 All permutations in $\mathcal{S}_1$ 
leave the spin factor $B_1$ invariant.  The 
integrals in eq.~\eqref{eq:w1-1} are shown
in fig.~\ref{fig:1loop-integrals}.  The factors of $\tfrac 1 2$ in the
summand in eq.~\eqref{eq:w1-1} are symmetry factors needed to
compensate for double counting in the summation over 
$\mathcal{S}_1$. The expressions \eqref{eq:1loop-pppmmm}
and \eqref{eq:w1-1} hold only through order
$\mathcal{O}(\epsilon^0)$. At ${\cal O}(\epsilon)$ eq.~\eqref{eq:w1-1}
receives contributions 
from additional integrals while equation \eqref{eq:1loop-pppmmm}
receives contributions from additional spin factors. The terms of higher 
order in $\eps$ have been computed only recently~\cite{SD}.

The other two scalar functions, $W_2^{(1)}$ and $W_3^{(1)}$, may be obtained 
from eq.~\eqref{eq:w1-1} by replacing the set of permutations
$\mathcal{S}_1$  by the sets $\mathcal{S}_2$ and 
$\mathcal{S}_3$, respectively, where 
\begin{align}
  \label{eq:perm2}
  \mathcal{S}_2 &= \lbrace (234561), (432165), (561234),
  (165432)\rbrace\,,\\
  \label{eq:perm3}
  \mathcal{S}_3 &= \lbrace (345612), (543216), (612345),
  (216543)\rbrace\,.
\end{align}  
The elements of each of the permutations sets $\mathcal{S}_1$,
$\mathcal{S}_2$ and $\mathcal{S}_3$ leave invariant the spin factors
$B_1$, $B_2$ and $B_3$, respectively.  The union of these three
permutations sets, $\mathcal{S}_0 = \mathcal{S}_1 \cup \mathcal{S}_2 \cup
\mathcal{S}_3$, is the set of all cyclic 
permutations and their reflections;
the MHV amplitude can be expressed
as a sum over this larger set.

The one-loop scattering amplitudes for the other two independent
helicity configurations have a structure similar to
eq.~\eqref{eq:1loop-pppmmm}; the scalar functions $W_i^{(1)}$ are
unchanged while the spin factors $B_1$, $B_2$ and $B_3$ are replaced
\cite{Bern:1994cg} by new spin factors $D_1$, $D_2$ and $D_3$ for
the $({+}{+}{-}{+}{-}{-})$ helicity configuration, and by $G_1$, $G_2$ and $G_3$
for the $({+}{-}{+}{-}{+}{-})$ configuration\footnote{The notable
  difference between $B_i$ and the non-split helicity spin factors
  $D_i$, $G_i$ is that, while the former are rational functions of
  products of adjacent spinors, the latter also contain products of
  non-adjacent spinors. This obscures their transformation
properties under the dual conformal symmetry~\cite{Drummond:2008cr},
which become manifest only when the amplitudes are combined
into a superamplitude~\cite{Drummond:2008vq}.}:
\begin{align}
  \label{eq:1loop-ppmpmm}
  A_6^{(1),\text{NMHV}}({+}{+}{-}{+}{-}{-}) &= \frac 1 2 \left(D_1
    W_1^{(1)} + D_2 W_2^{(1)} + D_3 W_3^{(1)}\right)\,,\\
  \label{eq:1loop-pmpmpm}
  A_6^{(1),\text{NMHV}}({+}{-}{+}{-}{+}{-}) &= \frac 1 2 \left(G_1
    W_1^{(1)} + G_2 W_2^{(1)} + G_3 W_3^{(1)}\right)\,.
\end{align}  
Infrared consistency then implies that the tree-level 
amplitudes for the corresponding helicity configurations 
are~\cite{Bern:1994cg},
\begin{align}
  \label{eq:tree-pppmmm}
  A_6^{(0),\text{NMHV}}({+}{+}{+}{-}{-}{-}) &= \frac 1 2 \left(B_1 +
    B_2 + B_3\right)\,,\\
  \label{eq:tree-ppmpmm}
  A_6^{(0),\text{NMHV}}({+}{+}{-}{+}{-}{-}) &= \frac 1 2 \left(D_1 +
    D_2 + D_3\right)\,,\\
  \label{eq:tree-pmpmpm}
  A_6^{(0),\text{NMHV}}({+}{-}{+}{-}{+}{-}) &= \frac 1 2 \left(G_1 +
    G_2 + G_3\right)\,.
\end{align}
The classic expression for these amplitudes was derived in ref.~\cite{MPX}.
In later sections we will see that the structure present in 
eqs.~\eqref{eq:1loop-pppmmm}, \eqref{eq:1loop-ppmpmm} 
and~\eqref{eq:1loop-pmpmpm} ---
in which only the spin factors change between various helicity
configurations of the external lines --- persists at higher loops as well.

\subsection{Superspace and Superamplitudes \label{ssec:superspace} }

On-shell superspace provides a very convenient way of organizing
amplitudes in \NeqFour~SYM theory and making manifest supersymmetry
relations between them.  The bosonic part of this superspace is
parametrized by the usual bosonic spinor variables
$\lambda_i,\tlambda_i$, related to the external momenta $k_i$ by
$k_{i \alpha \dot{\alpha}} = \lambda_{i \alpha}
\tlambda_{i \dot{\alpha}}$.  The fermionic part is parametrized by
Grassmann coordinates $\eta^A_i$, where $A=1, \cdots, 4$ is an
$R$-symmetry index.  The on-shell fields of the $\mathcal{N}=4$ theory
are assembled into a superfield,
\begin{equation}
\Phi(\eta)=
   g_-
+\eta^A\psi_A
+\frac{1}{2!}\eta^{A}\eta^{B}\phi_{AB}
+\frac{1}{3!}\eta^{A}\eta^{B}\eta^{C}\epsilon_{ABCD}\psi^{D}
+\frac{1}{4!}\eta^{A}\eta^{B}\eta^{C}\eta^{D} \epsilon_{ABCD}g_+~~.
\label{eq:superfield}
\end{equation}

A superamplitude is a generating function for the scattering
amplitudes of component fields, which may be identified as the
coefficients of the appropriate combinations of $\eta_i$
variables. 

 The component
amplitudes may be extracted by multiplying the superamplitude with the
appropriate superfield and integrating over all Grassmann
parameters: 
\be 
A_n(k_1, h_1; \dotsc; k_n, h_n)=\int
\prod_{i=1}^nd^4\eta_i \;\prod \Phi_{h_i}(\eta)\;\sA_n(k_1, \eta_1,
\dotsc, k_n, \eta_n)\,.  
\ee 
The superfields $\Phi_{h_i}(\eta)$ have
a single nonvanishing term corresponding to the field with helicity
$h_i$. 
\iffalse
Since the half supersymmetry manifest in this on-shell
superspace can be preserved at all stages of scattering amplitude
calculations, the equation eq.~\eqref{eq:components} holds to all
orders in perturbation theory.
\fi

As an example, the $n$-point NMHV gluon scattering amplitudes
appear inside the superamplitude as follows:
\begin{eqnarray}
  \label{eq:components}
  \sA_n(k_1, \eta_1, \dotsc, k_n, \eta_n) = \cdots &&+ \eta_1^4 \eta_2^4
  \eta_3^4 A_n(-,-,-,+,+,\dotsc,+) 
  \\
  &&+ \eta_1^4 \eta_2^4 \eta_4^4 A_n(-,-,+,-,+,\dotsc,+) + \cdots,
\nonumber
\end{eqnarray} 
where $\eta^4$ is the $SU(4)$-invariant expression $\tfrac 1 {4!}
\epsilon_{ABCD} \eta^A \eta^B \eta^C \eta^D$.  In extracting 
these component amplitudes, the $\eta$ variables corresponding to the 
positive-helicity gluons are supplied by the superfields~\eqref{eq:superfield}
while those for the negative-helicity ones appear explicitly
in the superamplitude. 
Because the half of the supersymmetries manifest in this on-shell
superspace can be preserved at all stages of scattering amplitude
calculations, eq.~\eqref{eq:components} holds to all
orders in perturbation theory.

\def\teta{\tilde\eta}
The dual superspace in which the superfield is given by,
\begin{equation}
{\widetilde \Phi}(\teta)=
   g_+
+\teta_A\psi^A
+\frac{1}{2!}\teta_{A}\teta_{B}\phi^{AB}
+\frac{1}{3!}\teta_{A}\teta_{B}\teta_{C}\epsilon^{ABCD}\psi_{D}
+\frac{1}{4!}\teta_{A}\teta_{B}\teta_{C}\teta_{D} \epsilon^{ABCD}g_-\,,
\end{equation}
has also been used, see for example refs.~\cite{FreedmanTree,
  Drummond:2008bq, Bern:2009xq,Drummond:2008vq}. While the expression
for the superamplitude is unchanged, component amplitudes are extracted
by differentiating with respect to selected superspace coordinates,
eight for MHV amplitudes, twelve for NMHV ones, etc.:
\be 
A_n(k_1, h_1; \dotsc; k_n, h_n)=
\prod {\widetilde\Phi}_{h_i}\biggl(\frac{\partial}{\partial\eta}\biggr)\;
\sA_n(k_1, \eta_1,
\dotsc, k_n, \eta_n)\,.  
\ee 
For pure-gluon amplitudes, the differentiation is solely with respect
to the Grassmann coordinates of the negative-helicity gluons.  The
structure of superfields is, however, unimportant for the computation
of superamplitudes.

In general, $n$-point tree-level scattering amplitudes can be written as
follows~\cite{Drummond:2008vq},
\begin{equation}
  \label{eq:superspace-tree}
\sA_n^{(0)} = 
%
%g^{n-2}
%
\frac {i\,\delt4 (\sum_{i=1}^n \lambda_i
  \tlambda_i) \delt8 (\sum_{i=1}^n \lambda_i \eta^A_i)}{\langle
  1 2\rangle \langle 2 3\rangle \cdots \langle n 1\rangle} \sum_{k=0}^{n-4}
\mathcal{P}_n^k,
\end{equation} 
where $\mathcal{P}_n^k$ are polynomials in the Grassmann variables
$\eta_i$ of degree $4 k$.  Invariance under $R$-symmetry implies that
$\mathcal{P}_n^k$ are invariant under $SU(4)$ rotations of the
Grassmann variables $\eta^A$.  The lowest-order term in the $\eta$
expansion has Grassmann weight $8$, while the highest-order term has
Grassmann weight $4 n - 8$.  CPT conjugation exchanges weight $4
k+8$ with weight $4 n - 4 k - 8$.  The $k=0$ term in
eq.~\eqref{eq:superspace-tree} has $\mathcal{P}_n^0 = 1$ and contains
all the MHV amplitudes. The NMHV amplitudes are contained in the $k=1$
term.  Similarly to equation \eqref{eq:components} and for the same
reason, equation \eqref{eq:superspace-tree} is expected to hold to all
orders in perturbation theory.  Higher-order corrections 
\iffalse
are expected to 
\fi
can alter only the coefficients of the polynomials $\mathcal{P}_n^k$,
{\it i.e.\/} the component amplitudes.   Throughout the paper,
four-fold bosonic momentum-conserving delta functions will appear,
products of delta functions over the four components whose indices
(a vector index $\mu$ or a pair $(\alpha,\dot\alpha)$ of spinorial
indices) we suppress.  A variety of four-fold Grassmann delta functions,
products of delta functions
taken over the $SU(4)$ index $A$, and eight-fold Grassmann delta functions,
products of delta functions taken over a pair of a spinor index $\alpha$
and an $SU(4)$ index $A$, will also appear.  In these delta functions,
we will suppress the (bosonic) spinor index, but display the (Grassmann)
$SU(4)$ index explicitly.

The tree-level MHV superamplitude was written down long ago
by Nair~\cite{Nair:1988bq},
\begin{equation}
\label{eq:MHV0}
  \sA_n^{(0),\text{MHV}} =\frac {i\,\delt4\left(\sum_{i=1}^n \lambda_i
      \tlambda_i\right) \delt8\left(\sum_{i=1}^n \lambda_i
      \eta^A_i\right)}{\langle 1 2\rangle \cdots \langle n 1\rangle}\,.
\end{equation}  
The $\overline{\text{MHV}}$ amplitude has an equally simple form in the 
conjugate superspace, whose coordinates are the conjugate spinors 
${\tilde \lambda}_i$ and the Fourier-conjugate ${\tilde\eta}$ of the Grassmann 
variables $\eta$. Fourier-transforming to the same superspace as the MHV 
amplitude implies~\cite{Drummond:2008bq} that the
 $\overline{\text{MHV}}$ superamplitude is 
\begin{equation}
\label{eq:MHVbar0}
  \sA_n^{(0),\overline{\text{MHV}}} = i\frac{\delt4 (\sum_{i=1}^n \lambda_i
  \tlambda_i)}{[1 2] \cdots [n 1]} \int
  d^8\omega \prod_{i=1}^n \delt4(\eta^A_i -
  {\tilde\lambda}^{\dot{\alpha}}_i \omega^A_{\dot{\alpha}}).
\end{equation}

Manifestly supersymmetric expressions for non-MHV amplitudes could be
obtained~\cite{GGK} through a supersymmetric generalization of the MHV
vertex expansion~\cite{FreedmanProof, CSW_LagrangianSusy,
  Kiermaier}. The expressions obtained this way do not {\it a priori\/}
exhibit any special properties. 
DHKS presented~\cite{Drummond:2008vq} a special form 
for $\mathcal{P}_6^1$, and showed that it enjoys
an extended symmetry, so-called dual superconformal symmetry.
Explicit expressions for all the $\mathcal{P}_n^k$ polynomials were
given by Drummond and Henn~\cite{Drummond:2008cr}, using a supersymmetric 
form~\cite{Brandhuber:2008pf, ArkaniHamed:2008gz} of the 
Britto, Cachazo, Feng, and Witten (BCFW)
on-shell recursion relations~\cite{OnShell}.

On-shell superspace encodes the relations between amplitudes that are
implied by supersymmetry, but does not identify the basic, irreducible
components from which all others can be obtained.  Identifying such
basic amplitudes, from which all others can be obtained via
supersymmetry transformations (along with the required sequence of
transformations) is in general a difficult problem.  Not all
corrections to the coefficients in the polynomials $\mathcal{P}_n^k$
are independent; as these coefficients are nothing but
component amplitudes, they are related by supersymmetry Ward
identities. Elvang, Freedman, and Kiermaier have provided
a solution~\cite{Elvang:2009wd} to this class of questions.

Apart from clarifying the structure of tree-level amplitudes,
knowledge of tree-level superamplitudes allows us to perform
manifestly-supercovariant higher-loop calculations using generalized
unitarity.  The one-loop calculation of ref.~\cite{Drummond:2008vq}
generalizes the result of ref.~\cite{Bern:1994cg} for the NMHV six-gluon
amplitudes to a manifestly supersymmetric expression encompassing all
possible external states. In
refs.~\cite{FreedmanDualConformal,Bern:2009xq} superamplitudes were
used to evaluate the sum over all the particles crossing generalized
unitarity cuts for $n$-point MHV amplitudes at any loop order.
In the section~\ref{TheCalculation} we will describe in detail the
steps needed for evaluating two- and higher-loop
superamplitudes for any number of external legs and Grassmann weight,
and elucidate the subtleties that arise in such evaluations.

\subsection{The six-point NMHV superamplitude}
\label{SixPointNMHVSuperamplitudeSection}

As our focus in later sections will be on the two- and higher-loop
six-point NMHV (super)amplitude, we first review and extend the
supersymmetric results of ref.~\cite{Drummond:2008vq} for the
tree-level and one-loop expressions for this amplitude.

As is true for the component amplitudes, relations between rational
functions of bosonic spinor products and Grassmann variables allow
the tree-level superamplitude to be expressed in several equivalent
forms. 
We may identify a canonical form, which will also be useful for 
higher-loop 
calculations, by starting from the $\eps$-pole terms in the 
one-loop superamplitude.  This superamplitude
is given by the supersymmetrization~\cite{Drummond:2008vq} of eqs.~\eqref{eq:1loop-pppmmm}, 
\eqref{eq:1loop-ppmpmm} and
\eqref{eq:1loop-pmpmpm},
\begin{equation}
  \label{eq:nmhv-1loop}
  \sA_6^{(1),\text{NMHV}} = \frac{a}{2} \sA_6^{(0),\text{MHV}}  \bigl(
    (R_{413} + R_{146}) W_1^{(1)} + (R_{524} + R_{251}) W_2^{(1)} +
    (R_{635} + R_{362}) W_3^{(1)} +
    \Ord(\eps)\bigr)\,,
\end{equation} 
where $A_6^{(0),\text{MHV}}$ is the tree-level MHV superamplitude, 
the loop expansion parameter $a$ is defined in eq.~\eqref{eq:loop-exp-param}
and the products $A^{(0),\text{MHV}} R_{j,j+3, j+5}$ with $j=1,\dots, 6\,$
(all indices understood mod~6) are,
\begin{eqnarray}
&&  \sA^{(0),\text{MHV}} R_{j,j+3,j+5} =\nonumber\\
&&\hskip 5mm \frac{%
    \delt8\bigl(\sum \lambda_i \eta^A_i\bigr)
    }
 {\spa{j}.{(j\tp 1)} \spa{(j\tp 1)}.{(j\tp 2)}\spb{(j\tp 3)}.{(j\tp 4)}
  \spb{(j\tp 4)}.{(j\tp 5)}}\\
&&\hskip 5mm \times\frac{\delt4\bigl(\eta^A_{j+3} \spb{(j\tp4)}.{(j\tp5)} 
           + \eta^A_{j+4} \spb{(j\tp5)}.{(j\tp3)} 
           + \eta^A_{j+5} \spb{(j\tp3)}.{(j\tp4)}\bigr)}
 {\sandmx{j}.{K_{j+1,j+2}}.{(j\tp 3)}
  \sandmx{(j\tp 2)}.{K_{j+3,j+4}}.{(j\tp 5)}
   s_{j,j+1,j+2}}\,.\nonumber
\end{eqnarray} 
This product is covariant under dual inversion, with the same
weight as the tree-level MHV superamplitude. 
For generic momentum configurations (that is, away from
soft or collinear configurations), the superfunctions
$R_{j,j+3,j+5}$ are thus invariant
under dual superconformal transformations.

The functions $W_i^{(1)}$ have identical poles in the dimensional
regularization parameter $\epsilon$; this reflects the
universality of infrared divergences.  A
canonical expression for the tree-level six-point NMHV superamplitude
is then simply,
\begin{equation}
  \label{eq:nmhv-tree}
  \sA_6^{(0),\text{NMHV}} = \frac{1}{2} \sA_6^{(0),\text{MHV}} \big(
    R_{146} + R_{251} + R_{362} + R_{413} + R_{524} + R_{635}
  \big )\,.
\end{equation} 
The $R$ invariants are not all independent; in the presence of the
super-momentum conservation constraint they obey the linear six-term
relation,
\begin{equation}
  \label{eq:R-identity}
  \sA_6^{(0),\text{MHV}} \big( R_{146} - R_{251} + R_{362} - R_{413}
    + R_{524} - R_{635} \big ) = 0~~.
\end{equation}
This relation, akin to relations derived from the Grassmannian formulation of
tree-level amplitudes~\cite{ArkaniHamed:2009dn},  leads to two apparently 
different presentations of the six-point NMHV superamplitude:
\begin{equation}
  \label{eq:nmhv-tree-alt}
  \sA_6^{(0),\text{NMHV}} 
  = \sA_6^{(0),\text{MHV}} \big( R_{146} +
    R_{362} + R_{524} \big)
  = \sA_6^{(0),\text{MHV}} \big(
    R_{251} + R_{413} + R_{635} \big)~~.
\end{equation}  
A proof of eq.~\eqref{eq:R-identity} amounts to showing that the first
expression in eq.~\eqref{eq:nmhv-tree-alt} can be derived from the
BCFW recursion relation~\cite{OnShell} with a supersymmetric
shift~\cite{ArkaniHamed:2008gz} while the second expression follows
from the cyclic symmetry of the superamplitude.
At higher loops the identity~\eqref{eq:R-identity} is crucial for ensuring 
the consistency of unitarity cuts.
It should also play a role in reconstructing scattering amplitudes from their 
leading singularities~\cite{ArkaniHamed:2009dn}. 

The six-point NMHV amplitude is special among NMHV amplitudes as it
exhibits a discrete invariance related to parity transformations. We
will discuss this symmetry and its consequences in the following.  A
similar discussion generalizes to the $2 n$-point N$^{n-2}$MHV
amplitudes.  As mentioned previously, (CPT) conjugation of
superamplitudes amounts to Fourier-transforming the Grassmann
coordinates (reversing the helicities of all
component fields) and exchanging spinors and conjugate spinors,
$\lambda_i \leftrightarrow \tlambda_i$.  It is easy to check that this
sequence of transformations maps the products $\sA_6^{(0),\text{MHV}}
R_{ijk}$ into themselves up to a cyclic permutation by three units:
\begin{equation}
\label{Rconjugation}
  \sA_6^{(0),\text{MHV}} R_{146} \to \sA_6^{(0),\text{MHV}} R_{413}, 
  \quad \text{etc.}
\end{equation}
This is the supersymmetric generalization of an obvious invariance of
the six-gluon NMHV scattering amplitudes.
Invariance of the six-point superamplitude under this transformation
in turn requires that the functions $W_i^{(1)}$ in equation
\eqref{eq:nmhv-1loop} be invariant under conjugation.

Apart from terms proportional to the sum of $R$ invariants, the
$\mathcal{O}(\epsilon)$ part of the one-loop amplitude also contain
terms which are proportional to differences of $R$ invariants. They
have been computed directly in a one-loop calculation~\cite{SD},
and their existence may also be inferred from the two-loop calculation we will 
describe in later sections.
% the next section.
%
As such differences are odd under conjugation, they must be accompanied 
by parity-odd (pseudo-scalar) functions $\widetilde{W}_i^{(1)}$.  
%%
%We will discuss these $\mathcal{O}(\epsilon)$ contributions in more 
%detail later. {\bf (adjustment needed)}
%%%%%%%%%%

\subsection{Dual Conformal Invariance and the Six-Point Superamplitude}

As mentioned above, DHKS showed~\cite{Drummond:2008vq} that tree-level
amplitudes are covariant, with weight $(-1)$, under dual
superconformal symmetry.  This property extends to the rational
functions $\sA_6^{(0),\text{MHV}}R_{ijk}$.  To what extent does the
symmetry extend to the full one-loop amplitude?

The dual conformal and dual superconformal symmetries are only defined
in four dimensions.  One possible extension is the notion of
pseudo-conformality: were we to regulate the integral functions
$W_i^{(1)}$ by off-shell continuation, they would become dual
conformal invariant, as they are sums of box integrals with the appropriate prefactors.
\iffalse
rendering them dimensionless.  
\fi
Additional evidence towards a kind of
dual conformal invariance comes from the
observation~\cite{Bullimore:2009cb,Korchemsky:2010ut} that leading
singularities are dual conformal invariant.

We can do better than this.  DHKS noticed~\cite{Drummond:2008vq} that
the ratio\footnote{This ratio is sensible because in chiral on-shell
  superspace any superamplitude is proportional to the super-momentum
  conservation constraint $\delt8(\sum_{i=1}^n \lambda_i \eta_i^A)$,
  which contains the entire Grassmann-dependent factor in the MHV
  amplitude.}  of the six-point NMHV to MHV superamplitudes, each
taken through one-loop order, is invariant under dual conformal
transformations.  That is, the ratio is invariant under
transformations that preserve the cross-ratios
\begin{equation}
  \label{eq:cross-ratios}
  u_1 = \frac{s_{12} s_{45}}{s_{123} s_{345}}, \qquad 
  u_2 = \frac{s_{23} s_{56}}{s_{234} s_{123}}, \qquad 
  u_3 = \frac{s_{34} s_{61}}{s_{345} s_{234}}\,.
\end{equation} 
The ratio of superamplitudes is a natural quantity, as it is
infrared finite.

In gauge theories, the structure of infrared divergences in
dimensional regularization is independent of the helicity
configuration~\cite{KnownIR,KorchemskyMarchesini}.  At one loop, for
example, the pole terms are proportional to the tree amplitudes.  This
makes the ratio of any helicity amplitude to the MHV amplitude
infrared finite.

The finiteness of such ratios makes it possible to take the
four-dimensional limit, and to inquire about their properties under
dual (super)conformal transformations.  Of course, finiteness does not
guarantee dual conformal invariance.  Indeed, the relation between
these two properties has been investigated in
ref.~\cite{Drummond:2008bq} with the conclusion that, in dimensional
regularization, there exist infrared-finite combinations of
pseudo-conformal integrals which are not dual conformal invariant.

\def\Coeff{C}
\def\Coefftilde{{\widetilde{C}}}
Explicit calculations show that such subtleties do not arise here and,
through one-loop order, the six-point NMHV superamplitude has the
factorized form~\cite{Drummond:2008vq}
\begin{equation}
  \label{eq:one-loop-nmhv}
  \sA_6^{\text{NMHV}} = \frac{1}{2}  \sA_6^{\text{MHV}} \left[ R_{146}
    \left(1 + a \Coeff_{146}^{(1)}\right) +
    \text{cyclic} + \mathcal{O}(a^2)\right]\,,
\end{equation}
with the functions $\Coeff_{i,i+3,i+5}$ manifestly expressed in terms
of the dual conformal ratios \eqref{eq:cross-ratios}:
\begin{equation}
  \label{eq:v1}
  \Coeff_{146}^{(1)} = - \ln u_1 \ln u_2 + \frac 1 2 \sum_{k=1}^3 \left(\ln
    u_k \ln u_{k+1} + \li_2 (1-u_k)\right) - \frac{\pi^2}{6} + \Ord(\eps)\,,
~~~~ \text{etc.}
\end{equation}
This function differs from $V^{(1)}$ as defined in ref.~\cite{Drummond:2008vq}
by $-\pi^2/6$, due to differences in normalization of amplitudes
and finite differences between the Wilson loop expression
and the one-loop amplitude.  It also differs in including $\Ord(\eps)$ terms; its 
$\eps$-independent part agrees with the function $V^{(1)}$ defined in ref.~\cite{Drummond:2008bq}.

For completeness we record~\cite{NeqFourOneLoop} the integral
representation of the one-loop six-point MHV amplitude through
$\Ord(\epsilon^0)$:
\begin{eqnarray}
\sA_6^{(1),\text{MHV}}&=&\sA_6^{(0),\text{MHV}}M^{(1)}_6,
\\
M^{(1)}_6&=&-\frac{1}{8}\sum_{\sigma\in \mathcal{S}_1 \cup \mathcal{S}_2 \cup \mathcal{S}_3}
\left(s_{12}s_{23}I^{\text{1m}}(\sigma)
    +\frac{1}{2}(s_{234}s_{345}  - s_{61} s_{34})I^{\text{2me}}(\sigma)\right)\,.
\nonumber
\end{eqnarray}

  In writing eq.~\eqref{eq:v1} we used the convention
that $u_{i+3} = u_i$.  
The ratio function is thus manifestly dual conformal invariant
through one loop.  It does not have the full dual superconformal invariance, 
dual supersymmetry being broken by a holomorphic anomaly~\cite{Korchemsky:2009hm}.

DHKS conjectured~\cite{Drummond:2008vq} that the main features of
eq.~\eqref{eq:one-loop-nmhv} survive higher-loop corrections: that
the six-point NMHV superamplitude may be factorized as
\begin{equation}
  \label{eq:all-loop-nmhv}
  \sA_6^{\text{NMHV}} =\frac{1}{2}  \sA_6^{\text{MHV}} \left[ R_6^{\text{NMHV}}
  + \mathcal{O}(\epsilon)\right]\,;
\end{equation}
and that the functions $R_6^{\text{NMHV}}$ have no further $\epsilon$
dependence, are well-defined in four dimensions and, to all loop
orders, are dual conformal invariant. The conjecture does not
specify the structure of the ${\cal O}(\epsilon)$ terms or of the spin
factors that enter the functions $R_6^{\text{NMHV}}$ beyond one-loop
level.
At one-loop, the $\mathcal{O}(\epsilon)$ terms are irrelevant to any
`physical' quantity.  However, these terms will contribute
nontrivially to both the divergent and finite parts of the 
$\Ord(a^2)$ terms in the product on the right-hand side of 
eq.~\eqref{eq:all-loop-nmhv}.
Our calculation will clarify the meaning of these one-loop terms for
that part of the amplitude dependent on parity-even combinations of
$R$ invariants.  We will show that they are determined by the 
$\Ord(\eps)$ terms in the one-loop NMHV amplitude, which have
been calculated recently by Dixon and Schabinger~\cite{SD}.

Before proceeding to describe our calculation, we will discuss in the
next section the structure of our result as well as the expected
properties of the resummed six-point NMHV amplitude.

\def\dpc{double-pentagon}
\def\sqrc{flying-squirrel}
%%%%%%%%%%%%%%%%%%%%%%%%%%%%%%%%%%%%%%%%%%
%%%%%%%%%%%%%%%%%%%%%%%%%%%%%%
\begin{figure}[ht]
\centerline{\includegraphics[scale=0.9]{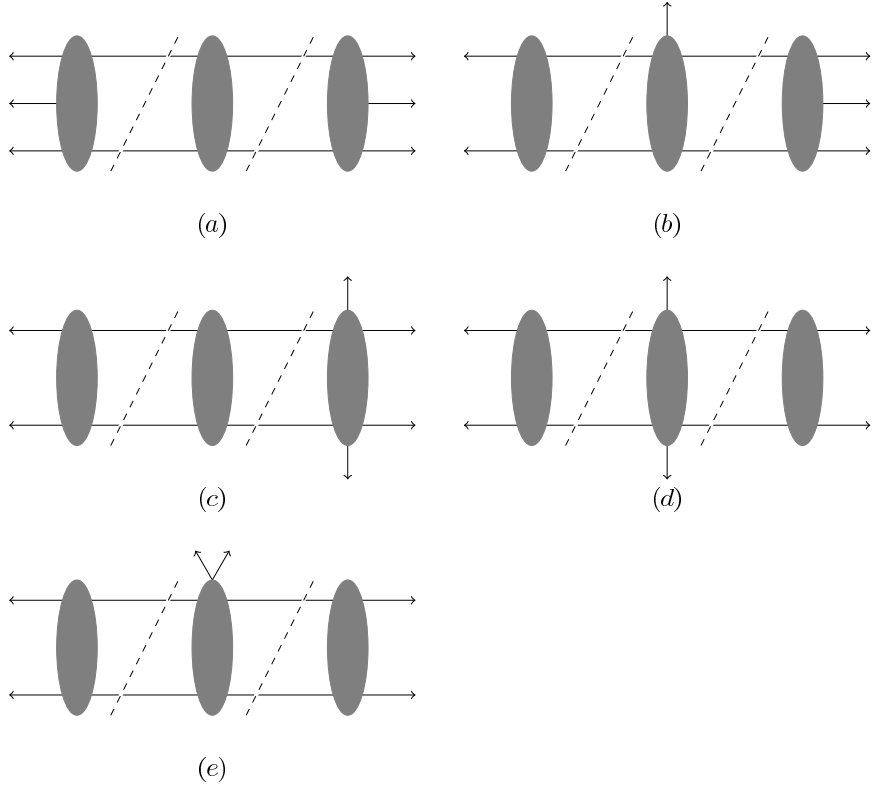}}
\caption{Generalized cuts required to determine the two-loop NMHV amplitude:
(a) the `\dpc' cut (b) the `turtle' cut (c) the `hexabox' cut
(d) the `\sqrc' cut (e) the `rabbit-ears' cut.
Unlike the MHV calculation, all permutations of the external legs 
must be considered.
\label{2loopcuts}}\nonumber
\end{figure}

\section{Structure of the Six-Point NMHV Amplitude \label{higher_loop_structure}}

In order to obtain the six-point NMHV amplitude to a given loop order,
we must determine all spin factors that appear, and construct the
functions of external momenta and coupling multiplying each one of
them. In the next section we will show explicitly that, through 
two-loop order and through $\Ord(\eps^0)$, the $R$ invariants are the only spin factors that
appear in the superamplitude.
The transformation of the $R$ invariants under conjugation (\ref{Rconjugation})
then implies that, through two-loop order, the superamplitude can be 
written as follows,
\begin{multline}
  \label{NMHVampAllLoop}
  \sA^{\text{NMHV}}_6 = 
  %
  %\frac{a^2}{8}
  %
  \frac{1}{2} \sA^{(0),\text{MHV}}_6
  \Bigl[ (R_{413} + R_{146}) W_1(a) + (R_{524} + R_{251})
  W_2(a) + (R_{635} + R_{362}) W_3(a)
  \\
 +(R_{413} - R_{146}) \widetilde{W}_1(a) + (R_{524} - R_{251})
 \widetilde{W}_2(a) + (R_{635} - R_{362}) \widetilde{W}_3(a) +
 \mathcal{O}(\epsilon) \Bigr]\,,
\end{multline}
where the $W_i(a)$ are scalar functions
and the $\widetilde{W}_i(a)$ are pseudoscalar functions.
We present the calculation of the two-loop
six-point NMHV superamplitude, computing explicitly the terms depending
on parity-even combinations of $R$ invariants.  
We will find that the four-dimensional cut-constructible part of
the parity-even functions $W^{(2)}_i$ can be expressed as a sum of
pseudo-conformal integrals.
We will also
confirm that, unlike their one-loop counterparts, the pseudoscalar
functions $\widetilde{W}^{(2)}_i$ have nonvanishing divergent and
finite parts in the $\eps$ expansion.  We will not compute these
functions explicitly, but the general infrared structure of
gauge theories
divergences requires that they have at most simple
($1/\eps$) poles, as both the tree and one-loop amplitudes
[through $\Ord(\eps^0)$] are free of such terms.
In this section, we describe the expected general structure
of the NMHV amplitude, and the structure of
its collinear limits.

\subsection{Beyond Two Loops}

We expect the pseudo-conformality of the coefficient functions
to continue to all loop orders.  To see this, consider a
four-dimensional generalized unitarity cut that decomposes an
$L$-loop superamplitude into a product of $k$ tree-level superamplitudes 
\be
\sA_n^{(L)}\Big|_{\text{cut}}=\prod \sA_{1}\dots \sA_{k}\,.
\ee 
As mentioned
earlier and shown in~\cite{Drummond:2008vq}, each superamplitude has
weight $(-1)$ under dual inversion.  Because a cut propagator simply
identifies the Grassmann variables and momenta of the legs that are
sewn, it has weight $(+2)$ under this transformation. Thus, the
product above together with the cut propagators has vanishing weight
for the cut legs and weight $(-1)$ for the external legs. This implies
that these cuts can all be saturated by cuts of pseudo-conformal
integrals.

Unlike the scalar functions $W_i^{(2)}$, 
the pseudo-scalar functions $\widetilde{W}^{(2)}_i$ are not uniquely
defined.  Indeed, the identity (\ref{eq:R-identity})
implies that it is possible to uniformly add 
 an arbitrary pseudoscalar function to the
$\widetilde{W}^{(2)}_i$ without
affecting the value of the amplitude. In particular, we could set any
one of these
functions to zero.  This ambiguity can be partly
eliminated by requiring that the superamplitude be manifestly
invariant under cyclic permutations of external legs:
\begin{equation}
  \label{eq:cyclic-sym}
  \widetilde{W}_i^{(2)} =
  \mathbb{P} \widetilde{W}_{i-1}^{(2)},
\end{equation} 
where $\mathbb{P}$ is the operation of permutation to
the right by one unit: 
\begin{equation}
  \mathbb{P} f[k_1,k_2,k_3,k_4,k_5,k_6] = f[k_2,k_3,k_4,k_5,k_6, k_1]\,.
%  ,\\
%  \mathbb{I} f[k_1,k_2,k_3,k_4,k_5,k_6] = f[k_6,k_5,k_4,k_3,k_2, k_1].
\label{shift_inv}
\end{equation}
%For later convenience we also defined an inversion operator.
The corresponding equation for the $W_i^{(2)}$ functions,
\begin{equation}
W_i^{(2)} = \mathbb{P} W_{i-1}^{(2)}\,,
\end{equation}
follows from the symmetry of the superamplitude.

Requiring cyclic symmetry does not completely fix the ambiguity
in the
pseudoscalar functions $\widetilde{W}^{(2)}_i$, as 
parity-odd cyclicly symmetric functions do exist. An example of such a
function is,
\begin{equation}
  f = \epsilon_{1234} f_1 + \epsilon_{2345} f_2 + \epsilon_{3456} f_3 +
  \epsilon_{4561} f_4 + \epsilon_{5612} f_5 + \epsilon_{6123} f_6,
\end{equation} 
where $f_i$ are parity-even functions of external momenta $k_j$ related by 
the action of the shift operator $f_i = \mathbb{P} f_{i-1}$ and 
$\epsilon_{ijmn} = \epsilon_{\mu \nu \rho \sigma} k_i^\mu
k_j^\nu k_m^\rho k_n^\sigma$.

  \begin{figure}
    \centering
    \includegraphics[scale=.5]{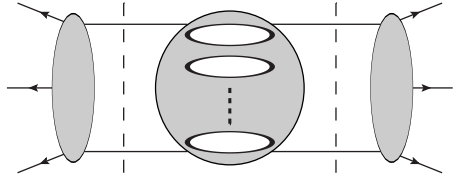}
    \caption{A cut of an $L$-loop six-point amplitude isolating an $(L-2)$-loop
    four-point amplitude with no external legs.  The cut is proportional
    to a lone $R$ invariant.}
    \label{fig:allloopcut}
  \end{figure}

The generalized-unitarity argument above does not reveal the
complete set of spin factors that appear at higher loops
in the six-point NMHV amplitude.
 The structure of leading singularities suggests~\cite{Bullimore:2009cb}
that new structures beyond the $R$ invariants
of one and two loops will be generated at three loops for
amplitudes with ten or more external legs, but 
that no new structures will appear beyond that order.  It also suggests
that no new invariants should appear beyond two loops for amplitudes
with seven or more, but fewer than ten, external legs; and that
no new invariants will appear beyond one loop for the six-point amplitude.

We can, however, argue that the spin factors present at one and two
loops will appear to all loop orders.  As we will see in the next
section, all tree-level $R$ invariants appear in double two-particle
cuts in a channel carrying a three-particle invariant.  Such a double
cut, shown in fig.~\ref{2loopcuts}(a), isolates a tree-level
four-point amplitude with no external legs attached to it.  An
all-loop generalization of this cut is shown in
fig.~\ref{fig:allloopcut}.  The well-known property of the four-point
amplitude at any loop order, that its spin factor is the same as at tree level,
implies that this cut will generate exactly the same spin
factors as at two loops. This argument extends trivially to all 
higher-loop contributions to the six-point amplitude that have
double two-particle cuts and isolate four- and five-point
amplitudes inside them.  It can be thought of
as a direct superspace generalization of the box
substitution rule~\cite{FiveLoop}.
At three loops and beyond, however, it is easy to construct cuts that are
outside this class.  Such cut-based arguments thus cannot rule
out spin factors beyond the $R$ invariants seen to date.

Apart from terms containing such new spin factors which may start
at three loops, the organization of the six-point NMHV amplitude in
(\ref{NMHVampAllLoop}) holds to all orders in perturbation theory. It
is therefore interesting to discuss the properties of the
parity-even and parity-odd functions,
\be
W_i(a)=1+a W_i^{(1)}+a^2 W_i^{(2)}+\dots ~~~\text{and}~~~~ {\widetilde
  W}_i(a)=1+a {\widetilde W}_i^{(1)}+a^2 {\widetilde W}_i^{(2)}+\dots~~
\label{Wfunctions}
\ee
in equation~(\ref{NMHVampAllLoop}) and of the finite functions 
\be
\Coeff_{i,i+3, i+5}(a)=1+a \Coeff_{i,i+3, i+5}^{(1)}+a^2 \Coeff_{i,i+3, i+5}^{(2)}+\dots
\ee
that appear in the ratio $A_6^{\text{NMHV}}/A_6^{\text{MHV}}$. 
The functions $\Coeff_{i,i+3, i+5}(a)$ will not have
definite parity. This is due to their relation to
linear combinations of functions with differing parity properties,
to wit
$(W_i(a)\pm {\widetilde W}_i(a))$, as well as to division by the
MHV amplitude which does not have definite parity properties.
For later convenience let us introduce 
the combinations $\Coeff_i(a)$ and $\Coefftilde_i(a)$,
\be
\label{new_ratios}
\frac{1}{2}(\Coeff_{i+3,i,i+2}+\Coeff_{i,i+3,i+5})\equiv \Coeff_i(a)&=&\frac{W_i(a)}{M_6(a)}\,,
\\
\frac{1}{2}(\Coeff_{i+3,i,i+2}-\Coeff_{i,i+3,i+5})\equiv \Coefftilde_i(a)&=&
\frac{{\widetilde W}_i(a)}{M_6(a)}\,.
\ee
The properties of $M_6(a)$ together with the universality of infrared
divergences implies, that through two loops, these functions have
definite parity up to corrections that vanish in the
$\epsilon\rightarrow 0$ limit\footnote{To guarantee that the functions
  $\Coeff_i(a)$ and $\Coefftilde_i(a)$ have definite parity to all
  orders in perturbation theory it is necessary to divide only by the
  parity-even part of $M_6(a)$.}\hskip -5pt.

\subsection{Collinear Limits}

The scalar and pseudo-scalar functions $W_i(a)$ and ${\widetilde
  W}_i(a)$ have specific properties dictated by the behavior of the
amplitude in collinear limits~\cite{AllOrdersCollinear}:
\be
A_6^{(L)}(\dots ,i^{\lambda_i},(i+1)^{\lambda_{i+1}},\dots)\rightarrow
\sum_{\lambda=\pm}\sum_{s=0}^L
\Split^{(s)}_{-\lambda}(z;i^{\lambda_i},(i+1)^{\lambda_{i+1}})
A_5^{(L-s)}(\dots ,k^\lambda,\dots)\,,~~~
\label{AmplitudeCollinearLimitI}
\ee
where $k=k_i+k_{i+1}$ and $z$ is the collinear momentum
fraction, $k_i\simeq zk$.  
We can rewrite this equation for the all-orders amplitude,
\be
A_6(\dots ,i^{\lambda_i},(i+1)^{\lambda_{i+1}},\dots)\rightarrow
\sum_{\lambda=\pm}
\Split_{-\lambda}(z;i^{\lambda_i},(i+1)^{\lambda_{i+1}})
A_5(\dots ,k^\lambda,\dots)\,.
\label{AmplitudeCollinearLimit}
\ee
The
properties of $\Coeff_i(a)$ and $\Coefftilde_i(a)$ are more intricate as
they involve additional contributions from $M_6(a)$.

We will find it easiest to discuss the collinear limits 
in components.
Because $W_i$ and ${\widetilde W}_i$ do not depend on the precise helicity
assignment to the external legs, it suffices to discuss the
split-helicity configuration.  In the three independent collinear
limits, the spin factors $B_i$ in eqs.~\eqref{eq:b1},
\eqref{eq:b2} and \eqref{eq:b3} behave as follows:
\be
1\parallel2:&& B_{1,3}\rightarrow{\rm Split}_-^{\rm tree}(1^+,2^+, k^-)
 A_5^{(0)}(k^+,3^+,4^-,5^-, 6^-)
         \,,~~~B_2 \rightarrow 0
\cr
5\parallel6:&& B_{1,2} \rightarrow {\rm Split}_+^{\rm tree}(5^-,6^-, k^+)
 A_5^{(0)}(1^+,2^+,3^+,4^-,k^-)
         \,,~~~B_3 \rightarrow 0
\cr
3\parallel4:&&B_{2,3} \rightarrow 
{\rm Split}_+^{\rm tree}(3^+,4^-, k^+)A_5^{(0)}(1^+,2^+,k^-,5^-,6^-) 
\cr
&&~~~~~~+ 
{\rm Split}_-^{\rm tree}(3^+,4^-, k^-)A_5^{(0)}(1^+,2^+,k^+,5^-,6^-)
         \,,~~~B_1 \rightarrow 0
\ee
The collinear limits of the parity-odd coefficients ${\widetilde
  B}_i$, which are contained in the parity-odd combinations of $R$
invariants, are similar except that the relative sign between the two
terms in the $3 \parallel 4$ limit is reversed.  Combining these limits
with the overall behavior of the amplitude~\eqref{AmplitudeCollinearLimit},
we find,
\be
\label{collinear}
1\parallel2:&& W_1+W_3\rightarrow r_-(z;1^+,2^+)\,
\IE(M_5^{\overline{\text{MHV}}}(k,3,4,5, 6))
\\
&& {\widetilde W}_1+{\widetilde W}_3\rightarrow r_-(z;1^+,2^+)\,
\IO(M_5^{\overline{\text{MHV}}}(k,3,4,5, 6))
\cr
5\parallel6:&& W_1+W_2\rightarrow r_+(z;1^-,2^-)\,
\IE(M_5^{\text{MHV}}(1,2,3,4,k))
\cr
&& {\widetilde W}_1+{\widetilde W}_2\rightarrow r_+(z;5^-,6^-)\,
\IO(M_5^{\text{MHV}}(1,2,3,4,k))
\cr
3\parallel4:&&W_2+W_3 \rightarrow 
  r_+(z;3^+,4^-)\, \IE(M_5^{\overline{\text{MHV}}}(1,2,k,5,6))
+r_-(z;3^+,4^-)\, \IE(M_5^{{\text{MHV}}}(1,2,k,5,6))
\cr
&&{\widetilde W}_2+{\widetilde W}_3 \rightarrow 
  r_+(z;3^+,4^-)\, \IO(M_5^{\overline{\text{MHV}}}(1,2,k,5,6))
- r_-(z;3^+,4^-)\, \IO(M_5^{{\text{MHV}}}(1,2,k,5,6))
\nonumber
\ee
in which,
\be
r_{-\lambda}(z;i^{\lambda_i},(i+1)^{\lambda_{i+1}})=
\frac{{\rm Split}_{-\lambda} (z;i^{\lambda_i},(i+1)^{\lambda_{i+1}})}
{{\rm Split}_-^{\rm tree} (z;i^{\lambda_i},(i+1)^{\lambda_{i+1}})}\,;
\label{rescaled_split}
\ee
$M_5^{\text{MHV}}$ and $M_5^{\overline{\text{MHV}}}$ are the ratios
of the resummed five-point MHV and $\overline{\text{MHV}}$
amplitudes to their tree-level counterparts; and $\IE$ and $\IO$
denote projection operators onto the parity-even and parity-odd
components.  Functions not explicitly mentioned are unconstrained.

The collinear properties of the functions $\Coeff_{i,i+3,i+5}$ and
$\Coeff_{i+3,i,i+2}$ can be easily found by combining equations
(\ref{new_ratios}), (\ref{collinear}) and the collinear properties of
the MHV ratio $M_6=A_6^{\text{MHV}}/A_6^{(0),\text{MHV}}$. In
particular, they contain the Levi-Civita tensors necessary to
transform, for example in the $1 \parallel 2$ limit, the MHV five-point
amplitude factor into an \MHVbar{} five-point amplitude.

The iteration relation for the rescaled splitting amplitude
(\ref{rescaled_split})~\cite{ABDK} suggests another natural
organization of the functions $W_i$, which is similar in spirit to the
BDS ansatz~\cite{BDS}:
\be
\ln W_i=\sum_{l=1}^\infty a^l\left[f_l(\epsilon)W_i^{(1)}(l\epsilon)
+C_l+R_{6;i}^{(l)}+\mathcal{O}(\epsilon)\right]~~.
\label{eq:Witeration}
\ee
The structure of infrared singularities and the collinear behavior
require that $\mathcal{O}(\epsilon^0)$ and $\mathcal{O}(\epsilon^1)$
terms in the functions $f_l(\epsilon)$ be the same as for the
six-point MHV amplitude.
The functions $R_{6;i}^{(l)}$, are similar in spirit to the
remainder function $R_6^{(l)}$ of the six-point MHV amplitude. They
are closely related to the functions $\Coeff_i(a)$ introduced in
equation~(\ref{new_ratios}):
\be
\Coeff_i(a)= 
\exp{\left[\gamma_K(a)\left (W_i^{(1)}-M_6^{(1)}\right)\right]}
\exp{\left[R_{6;i}(a)-R_{6}(a)\right]}+\mathcal{O}(\epsilon)\,,
\ee
where $\gamma_K(a)$ is the cusp anomalous dimension
%, $\Coeff_i^{(1)}=W_i^{(1)}-M_6^{(1)}$ 
and $R_{6;i}(a) = \sum_{l=2}^\infty a^l R_{6;i}^{(l)}$, etc. 
A natural consequence of the conjecture that $\Coeff_i(a)$ are invariant
under dual conformal transformations is that the remainder-like
functions $R_{6;i}(a)$ are also invariant. We will see that this is
indeed so.

The expectation~\cite{AM1, Abel:2007mw} that to leading order in the 
strong coupling limit, all amplitudes  with the same number of external legs 
are identical (or, equivalently, that 
$\lim_{a\rightarrow\infty} \ln\Coeff_i(a)=\Ord(a^{0})$ 
rather than $\Ord(\sqrt{a})$) 
predicts a simple relation between the remainder functions 
$R_{6;i}$ and the MHV remainder function $R_6$ to this order. 
Indeed, using the one-loop relation
\be
W_i^{(1)}-M^{(1)}=\Coeff_i^{(1)}
\ee
and the known value of the strong coupling expansion of the cusp anomaly 
\cite{Kruczenski, CuspStrongCoupling, BES, BKK}, it follows that 
\be
\frac{R_6-R_{6;i}}{\Coeff_i^{(1)}}=\frac{\sqrt{\lambda}}{\pi} \,,
\ee
with $\Coeff_i^{(1)}$ given in eq.~\eqref{eq:v1}. Using the numerical
results presented in later sections one may check that the 
weak-coupling expansion of the ratio appearing on the left-hand side
depends on the spin factor labeled by $i$; it seems therefore that a
relation of this type may hold only in the strong-coupling
limit.
%

%%%%%%%%%%%%%%%%%%%%%%%%%%%%%%%%%
\subsection{Triple-collinear limits}

Multi-collinear limits provide a richer set of constraints on
amplitudes with at least six external legs. Unlike the collinear
limits discussed in the previous section, they probe the detailed
structure of the dual-conformal invariant functions
unrelated to the infrared structure of the amplitude. 
In the case of the
six-point MHV amplitude, they provided a physical interpretation of the
remainder function~\cite{Bern:2008ap}. 
The most detailed limit we can consider with six external legs
involves three adjacent external
momenta becoming collinear,
\be
k_a=z_1 P\,,
\quad
k_b=z_2 P\,,
\quad
k_c=z_3 P\,,
\quad
z_1+z_2+z_3=1\,,
\quad 0\le z_i\le 1\,,
\quad
P^2\rightarrow 0\,.
\label{momentumfractions}
\ee
Let us understand what such limits
imply about the six-point NMHV amplitude and, in particular, about
the remainder-like functions $R_{6;i}(a)$.

An $L$-loop $n$-point amplitude factorizes as 
follows~\cite{AllOrdersCollinear}:
\be
A_n^{(L)}(k_1,\dots , k_{n-2},k_{n-1},k_{n})\mapsto
\sum_{\lambda=\pm}\sum_{s=0}^L
A_n^{(L-s)}(k_1,\dots ,P^{\lambda})\;
\Split^{(s)}_{-\lambda}(k_{n-2}k_{n-1}k_{n}; P)\,.
\label{all_loop_factorization}
\ee
Taking into account parity and reflection symmetries,
there are six independent triple-collinear splitting 
amplitudes~\cite{Bern:2008ap}:
\be
&&\Split_{+}(k_a^{+}k_b^{+}k_c^{+}; P),
~~~~~~~~
\label{lambdasum4}\\
&&\Split_{-\lambda_P}(k_a^{\lambda_a}k_b^{\lambda_b}k_c^{\lambda_c}; P),
~~~~~~~~
\lambda_a+\lambda_b+\lambda_c-\lambda_P=2\,,
\label{lambdasum2}\\
&&\Split_{-\lambda_P}(k_a^{\lambda_a}k_b^{\lambda_b}k_c^{\lambda_c}; P),
~~~~~~~~
\lambda_a+\lambda_b+\lambda_c-\lambda_P=0\,.
\label{lambdasum0}
\ee
The first one~(\ref{lambdasum4}) vanishes in any supersymmetric theory.
The three triple-collinear splitting amplitudes of the second 
type~(\ref{lambdasum2}), an example of which is 
$\lambda_a=\lambda_b=\lambda_c=\lambda_P=1$,
appear in limits of MHV amplitudes. 
The ${\cal N}=4$ supersymmetry Ward identities for MHV amplitudes imply
that their rescaled forms\footnote{%
We omit a trivial dimensional dependence on $s_{abc}$ from the
argument list of $r_S^{(l)}$.}
are all equal,
\be
\frac{\Split^{(l)}_\mp(k_a^\pm k_b^+k_c^+; P)}
{\Split^{(0)}_\mp(k_a^\pm k_b^+k_c^+; P^\mp)}=
\frac{\Split^{(l)}_\mp(k_a^+ k_b^\pm k_c^+; P)}
{\Split^{(0)}_\mp(k_a^+ k_b^\pm k_c^+; P)}=
r_S^{(l)}({\textstyle{\frac{s_{a b}}{s_{a b c}}}}, 
{\textstyle{\frac{s_{b c}}{s_{a b c}}}}, z_1, z_3)\,.
\label{RescaledSplitting}
\ee
These splitting amplitudes are relevant only for NMHV amplitudes with
at least seven external legs. They do not arise in the
factorization of six-point amplitudes, because
the four-point amplitude entering the factorization
(\ref{all_loop_factorization}) vanishes identically.

The two splitting amplitudes of the third kind~(\ref{lambdasum0}) arise 
only in limits of NMHV amplitudes and do not have a simple factorized
form similar to (\ref{RescaledSplitting})%
\footnote{The spin-averaged absolute values squared of tree-level
triple-collinear splitting amplitudes were computed in
ref.~\cite{Campbell}; without spin-averaging,
in refs.~\cite{CataniTripleCollinear}. The tree-level
triple (and higher) collinear splitting amplitudes themselves were
computed in ref.~\cite{Birthwright} using the MHV
rules~\cite{CSW}.  The one-loop correction to the 
$q\rightarrow q{\bar Q}Q$ triple-collinear splitting amplitude in QCD
was computed in ref.~\cite{CdFR}.}.
They are however the only splitting amplitudes that can appear in the 
triple-collinear limit of the six-point NMHV amplitude. 

As is true for the tree-level NMHV amplitudes, the splitting
amplitudes (\ref{lambdasum0}) have several different presentations
related by potentially nontrivial spinor identities.  A canonical one,
that is useful for our purpose, is obtained from the triple-collinear
limit of the six-point tree-level amplitude in
equations~(\ref{eq:tree-pppmmm})-(\ref{eq:tree-pmpmpm}).

%\newpage

As the functions $W_i$ are independent of the helicity assignment of
the external legs, we again discuss only the split helicity
configuration. Up to conjugation and relabeling the only non-trivial
limit is $2\parallel3\parallel4$. With the momentum fractions
$k_2=z_1 P,\, k_3=z_2 P,\,k_4=z_3 P$, the spin factors $B_i$ become:
\be
b_1&=&\frac{B_1}{A_4^{(0)}(1^+P^+5^-6^-)}\mapsto 
\frac{(1-z_3)^2}{\sqrt{z_1z_2z_3} \langle 23\rangle (\sqrt{z_1}[24]+\sqrt{z_2}[34])  }
\nonumber\\
b_2&=&
\frac{B_2}{A_4^{(0)}(1^+P^+5^-6^-)}\mapsto 
-\frac{(\sqrt{z_1}\langle24\rangle+\sqrt{z_2}\langle 34\rangle)^3}
{s_{234} \langle 23\rangle  \langle 34\rangle 
(\sqrt{z_2}\langle23\rangle+\sqrt{z_3}\langle 24\rangle) }
\cr
&&~~~~~~~~~~~~~~~~~~~~~~~~~~
+
\frac{[23]^3}{s_{234}[34](\sqrt{z_1}[24]+\sqrt{z_2}[34])(\sqrt{z_2}[23]+\sqrt{z_3}[24])  }
\\
b_3&=&\frac{B_3}{A_4^{(0)}(1^+P^+5^-6^-)}\mapsto 
\frac{z_2^{3/2}}{\sqrt{z_1 z_3}(1-z_1)[34](\sqrt{z_2}\langle23\rangle+\sqrt{z_3}\langle 24\rangle) }
\cr
&& ~~~~~~~~~~~~~~~~~~~~~~~~~~
+
\frac{(z_1 z_3)^{3/2}}{\sqrt{z_2}\langle 34\rangle (\sqrt{z_2}[23]+\sqrt{z_3}[24]) }~~.
\nonumber\ee
The tree-level splitting amplitude is simply
\be
\Split_{-}^{(0)}(k_2^+k_3^+k_4^-;P)=\frac{1}{2}(b_1+b_2+b_3)~~.
\ee
Thus, while these splitting amplitudes do not have a simple factorized
form similar to that for splitting amplitudes of the second 
type~(\ref{RescaledSplitting}), we see that the structure
of the six-point amplitude (\ref{NMHVampAllLoop}) implies that to this
order each component $b_i$ is dressed at higher loops by scalar
functions of momenta,
\be
\Split_{-}(k_2^+k_3^+k_4^-;P)=\frac{1}{2}\left(b_1w_1(a)+b_2w_2(a)+b_3w_3(a)\right)~~.
\ee
The parity-odd spin factors ${\widetilde B}_i$ also have nontrivial
triple-collinear limits.  Their coefficients 
${\widetilde W}_i$, though, must
contain Levi-Civita tensors and thus naively vanish in this limit.  The
triple-collinear limits of additional spin factors that may appear beyond
two loop order must be considered separately.

As was true for the limit discussed in ref.~\cite{Bern:2008ap}, none of
the conformal cross-ratios (\ref{eq:cross-ratios}) vanish as
$2\parallel3\parallel4$; they become
\be
{\bar u}_1=\frac{z_1z_3}{(1-z_1)(1-z_3)}\,,
\qquad
{\bar u}_2=\frac{s_{23}}{s_{234}}\,\frac{1}{1-z_3}\,,
\qquad
{\bar u}_3=\frac{s_{34}}{s_{234}}\,\frac{1}{1-z_1}\,.
\ee
Thanks to their expected dual conformal invariance (which we will
confirm in later sections), the remainder-like functions $R_{6;i}(a)$
retain their complete kinematic content, and may be read off the
two-loop triple-collinear splitting amplitude (\ref{lambdasum0}) by
subtracting the triple-collinear limit of the two-loop iteration of
the one-loop functions $W^{(1)}_i$.

%%%%%%%%%%%%%%%%%%%%%%%%%%%%%%%%%%%%%%%%%%%%
%%%%%%%%%%%%%%%%%%%%%%%%%%%%%%%%%%%%%%%%%%%%

\section{Constructing the Even Part of the Two-Loop Amplitude
\label{TheCalculation}}

We will construct the even part of the two-loop six-point NMHV
amplitude using 
a superspace form~\cite{Drummond:2008bq,Bern:2009xq}
of the generalized unitarity 
method~\cite{NeqFourOneLoop,Bern:1994cg,UnitarityII,
  NMHV7point,OneloopTwistorB,OneLoopNMHV,BCFUnitarity}.
On general grounds, the result will be
expressed as a sum of planar two-loop Feynman integrals with 
coefficients that are rational functions of the spinor variables.
 At this order, one-loop calculations suggest that
it is possible to exclude integrals with
triangle or bubble subintegrals.

Similarly to the two-loop
MHV amplitude, we will find that neither $W^{(2)}_i$ nor
$\widetilde{W}^{(2)}_i$ can be completely determined by
four-dimensional cuts.  Rather, they receive both divergent and finite
nontrivial contributions from integrals whose integrand is
proportional to the $(-2\epsilon)$ components of the loop momenta.  It
is quite nontrivial that these latter contributions can be organized
in terms of the same $R$ invariants as the four-dimensional
cut-constructible terms.

The generalized cuts that determine the amplitude are then the ones shown
in fig.~\ref{2loopcuts}, which are the same ones
that determine the MHV amplitude
\cite{Bern:2008ap}. Unlike the calculation of the MHV 
amplitude, however,
here it is necessary to evaluate cuts with all external helicity configurations,
as each yields information about different spin factors.

In any supersymmetric theory the improved power-counting ensures that
at one-loop order and through $\mathcal{O}(\epsilon^0)$ all terms can be
detected in four-dimensional cuts. Beyond one loop this is no
longer true generically; for example, the six-point MHV amplitude at
two loops receives nontrivial contributions from integrals whose
integrand vanishes identically when evaluated in four dimensions.
Four-point amplitudes in the \NeqFour~SYM theory are an exception:
through five loops they appear to be determined solely by four-dimensional
cuts.
We therefore decompose the functions $W_i^{(2)}$ in
eqs.~\eqref{NMHVampAllLoop} and \eqref{Wfunctions} into a
four-dimensional cut-constructible part and a part that requires
$D$-dimensional calculations,
\be
W_i^{(2)}=W_i^{(2),D=4}+W_i^{(2),\mu}~~.
\ee
For the former, powerful helicity and supersymmetry methods can be
employed. The latter part of the amplitude is determined by comparing
the result of $D$-dimensional and four-dimensional calculations and is
expressed in terms of ``$\mu$-integrals'' --- nontrivial integrals
whose integrand vanishes identically in four dimensions.

While all cuts may be evaluated easily, separating their contributions
to each one of the functions $W_i^{(2),D=4}$ is not always
straightforward.  As mentioned previously, (multiple) cuts in
channels carrying three-particle invariants capture a single even and odd spin factor at
a time and thus determine terms in a single $W_i^{(2),D=4}$ and
${\widetilde W}_i^{(2),D=4}$, with the index $i$ determined by the
helicity configuration of external legs. This is the case for cuts
({\it a\/}) and ({\it b\/}) in fig.~\ref{2loopcuts}.
In contrast, (multiple) cuts in channels carrying 
only two-particle invariants contribute 
simultaneously to several spin structures and thus to several  $W_i^{(2),D=4}$ 
and ${\widetilde W}_i^{(2),D=4}$ functions. 
This feature is already present in the cut construction of the
one-loop amplitude; in that case however, cuts in channels carrying
three-particle invariants suffice to completely determine the
amplitude~\cite{Bern:1994cg}.  The similarity between the expression
for cut ({\it c\/}) of fig.~\ref{2loopcuts}
 and a cut of the one-loop amplitude makes it possible to
disentangle it.
The cuts of fig.~\ref{2loopcuts}({\it d\/}) and ({\it e\/})
 however seem intractable in a component approach.

The component approach also fails to incorporate in a transparent way
the constraints imposed by supersymmetry.  On-shell superspace
provides the additional structure necessary for identifying the
contributions of the remaining cuts to each of the $W_i^{(2),D=4}$.
We shall therefore formulate the entire calculation of the
four-dimensional cut-constructible part of the amplitude in on-shell
superspace. After a brief overview of the structure of supercuts and
of the techniques necessary to disentangle them, we will discuss cut
({\it a\/}), and then proceed to a more detailed analysis of the
challenging cuts ({\it c\/}), ({\it d\/}) and ({\it e\/}).  For the
latter cuts we shall use a superspace generalization of the maximal
cut method \cite{FiveLoop, CompactThree}.

%%%%%%%%%%%%%%%%%%%%%%%%%%%%%%%%%%%%%%
\subsection{Unitarity in Superspace: General Features and Techniques\label{unitarity_general}}

Generalized cuts may be classified following the number of cut
conditions they impose.  The same is true for generalized supercuts.
At $L$ loops in four dimensions it is possible to impose at most $4L$
cut conditions; based on one- and two-loop information, it it likely that 
their solutions generically form a discrete set.  
This type of cut has been considered in the maximal-unitarity approach
as well as in the leading-singularity approach.
Maximal cuts, {\it i.e.} cuts with the maximal number of cut propagators, 
are typically insufficient to completely determine an
amplitude. For example, at two loops one frequently encounters double
box integrals, which cannot be detected by cutting eight
propagators. Near-maximal cuts, obtained by successively relaxing cut
condition in maximal cuts, provide an algorithmic way of identifying
these contributions.
Near-maximal cuts exhibit additional propagator-like singularities
which are exploited in the leading singularity approach to reduce the
one-parameter family of solutions to the cut conditions to a discrete
set.

The two-loop six-point NMHV amplitude can in principle be determined
entirely from the iterated two-particle cuts shown in
fig.~\ref{2loopcuts}. The Feynman integrals that contribute only to
cuts ({\it c\/}), ({\it d\/}) and ({\it e\/}) are also detected by certain
near-maximal cuts.  We have used them instead to check that the
resulting amplitude correctly reproduces cuts ({\it c\/}), 
({\it d\/}) and ({\it e\/}), supplemented by an additional cut
condition isolating terms in one of the tree amplitudes.

General supercuts are constructed~\cite{FreedmanTree, Drummond:2008bq,
  Bern:2009xq} by multiplying together superamplitudes, identifying
the $\eta$ parameters of the lines that are sewn together and
integrating over the common values of the internal $\eta$ variables.
The structure and properties of general supercuts have been analyzed
in detail in ref.~\cite{Bern:2009xq} where it was shown that, upon use
of a supersymmetric generalization of the MHV vertex rules~\cite{CSW},
their building blocks are generalized supercuts constructed only out of
 MHV and \MHVbar{} tree-level amplitudes.

When evaluating a supercut one encounters the situation that on one
side of the cut a momentum is outgoing and on the other side it is
incoming. In order to write the tree-level amplitude and in particular the
argument of their delta functions, it is necessary to define the
spinors $|{-}p\rangle$ and $|{-}p]$ corresponding to the incoming momentum
  $(-p)$. We use the analytic continuation rule~\cite{FreedmanTree}
 that the change in
  sign of the momentum is realized by a change of sign of the
  holomorphic spinor 
\be p\mapsto -p
  &\leftrightarrow& ~~~~~~\lambda_p\mapsto -\lambda_p\,,~~~~~~~~~~
  {\tilde\lambda}_p\mapsto +{\tilde \lambda}_p \,;\cr 
  &\leftrightarrow&
   ~~~|{-}p\rangle \mapsto -|p\rangle \,,~~~~~~~|{-}p]\mapsto|p]\,.
\ee

Let us discuss in detail the building blocks we require,
supercuts
constructed only out of MHV and \MHVbar{} tree-level amplitudes. 
Their evaluation requires the evaluation of
integrals of products of delta functions with arguments linear in Grassmann 
parameters, see eqs.~(\ref{eq:MHV0}) and (\ref{eq:MHVbar0}). 
For a $p$-particle cut of an N$^k$MHV amplitude this product contains $(8+4(k+p))$
delta function factors of which $(8+4k)$ remain upon integration. As  
discussed in refs.~\cite{FreedmanTree, Drummond:2008bq, Bern:2009xq}, 
the integration over the internal $\eta$ parameters realizes the sum over the states crossing 
the (generalized) supercut. 
For any supercut, eight of these delta functions can always be
singled out: they enforce the super-momentum conservation of the
amplitude,
\be
\label{eq:overall_delta}
\delt8(\sum_{i\in{\cal E}}\lambda_i\eta^A_i)
\ee
where ${\cal E}$ denotes the set of external lines.  These
delta functions may be thought
of as the supersymmetric generalization of the usual momentum
conservation constraint.
They may be extracted
without carrying out any Grassmann integrations, by taking suitable
linear combinations of the arguments of all delta functions. If the
supercut contains at least one MHV superamplitude factor, the Jacobian
of this transformation is unity. Their extraction also makes manifest
the invariance of the amplitude under half of the maximal
supersymmetry. Invariance under the other half of the supersymmetry,
generated by
\begin{equation}
  \bar{q}{}_A^{\dot \alpha} = \sum_{i=1}^n {\tlambda}^{\dot\alpha}_i 
  \frac {\partial}{\partial \eta^A_i}~~,
\end{equation} 
is not manifest, but can in principle be checked at the level of the 
Grassmann integrand.

The delta functions \eqref{eq:overall_delta} represent the complete
Grassmann parameter dependence of a supercut of an MHV amplitude.  The
Grassmann integrals simply yield the determinant of the system of
linear equations which are the arguments of the other $4p$ delta
functions, where $p$ is the number of cut lines~\cite{Bern:2009xq}.

For cuts of an N$^k$MHV amplitude there is a certain amount of freedom in evaluating 
the internal Grassmann integrals.
In general, however, the resulting $4k$ delta functions have many
undesirable features.  The essential ones are that (1) their arguments
may depend on loop momenta (if the cut conditions do not completely
freeze the momentum integrals) and (2) they may not make the
symmetries of the amplitude manifest.
We wish to express these Grassmann delta functions in terms of
structures that appeared at lower-loop order; in the case of the
six-point NMHV amplitude; these are the dual superconformal
$R$ invariants.  This is a non-trivial operation, and we have but
a limited set of tools available.

Given a set of $4k$ delta functions
\be
\prod_{i=1}^k\delt4(e_i(\eta, \lambda))\,,
\ee
it may be possible to construct linear combinations of their arguments 
$M_{ij}(\lambda)e_j(\eta,\lambda)$ which factorize into products of the 
desired combinations of spinors and Grassmann variables upon use of 
momentum and super-momentum conservation, cut conditions, and the fact that a 
Grassmann delta function equals its argument. For $k=1$, which is the case 
of interest to us, no linear  combinations can be constructed.

A possible strategy for eliminating the dependence of the delta
functions on loop momenta is to make use of the fact that a Grassmann
delta function equals its argument.
This observation replaces a cut carrying a Grassmann delta function
with a sum of cuts of tensor integrals with Grassmann-valued
coefficients. Albeit nontrivial due to their high rank, the tensor
integrals may then be reduced following the standard strategy of
integral reduction. While indeed successful in eliminating the loop
momentum dependence from the Grassmann delta functions, this strategy
is likely to lead to rather unwieldy expressions. We will not pursue
this direction.

An alternate approach to reorganizing Grassmann delta functions is to use the
Lagrange interpolation formula, which is most efficient when applied
to next-to-maximal cuts, which impose $(4L-1)$
on-shell conditions. Let $y$ be the variable that parametrizes the
solution to these cut conditions. The product of the $4k$ Grassmann
delta functions is then just a polynomial $P_d(y)$ of degree $d=4k$
with Grassmann-valued coefficients.
%
%For such polynomials, the Lagrange interpolation formula reads
Any such polynomial may be written as
\begin{equation}
  P_d(y) = \sum_{i=1}^{d+1} \prod_{\substack{j=1\\j \neq i}}^{d+1} 
  \frac{y-y_j}{y_i - y_j} P_d(y_i)\;,
\label{LIf}
\end{equation} 
where the values $y_i$ are arbitrary.
This equation simply encodes the fact that a polynomial of degree $d$
is determined by its values at $d+1$ points.

Choosing the points $y_i$ can be regarded as freezing the momentum
component unfixed by the cut condition; from this perspective it is
akin to the leading-singularity method which uses additional cut-like
conditions for the same purpose.
The Lagrange interpolation formula (\ref{LIf}) provides a different strategy, as
the points $y_i$ need not be chosen following the leading-singularity
prescription. If the two approaches are to agree, the residue of the
leading singularity must be proportional to the Grassmann delta
functions appearing in dual superconformal invariants. Evidence that
this is indeed true has been presented in
ref.~\cite{Bullimore:2009cb}. In general, however, in order to use the
interpolation formula (\ref{LIf}), there must exist more $y_i$ such
that $P_d(y_i)$ is (proportional to) a dual superconformal invariant
than are given by the leading-singularity approach.

In the next subsection we will use this strategy to analyze certain
seven-particle cuts of the six-point two-loop NMHV superamplitude. As we
will see, with judiciously chosen points $y_i$ it is possible to have
$P_4(y_i)$ be proportional to the delta functions appearing in the $R$
dual superconformal invariants.

Because of the arbitrariness in the choice of the $y_i$, the
decomposition of $P_d(y)$ in a linear combination of ``good''
Grassmann delta functions, such as the delta functions appearing in
the dual superconformal invariants, is not unique.  This signals the
existence of linear relations between the dual superconformal
invariants.  For six-point amplitudes, an identity arising this way is
eq.~\eqref{eq:R-identity}, which was already required for the consistency
of the various possible presentations of the tree-level amplitude. It
is conceivable that at higher points and/or higher loops, new
relations arise, beyond those that can be obtained from tree-level
considerations.

\begin{figure}[ht]
\centerline{\includegraphics[scale=0.8]{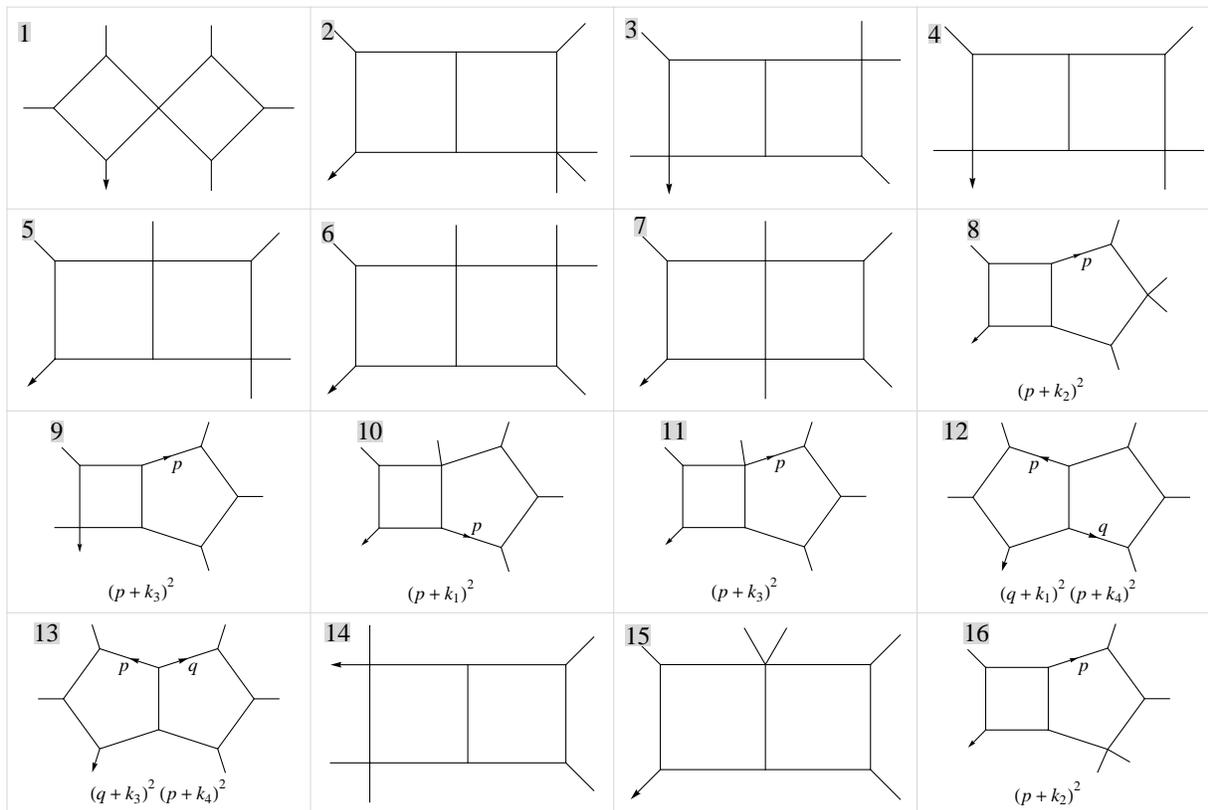}}
\caption{Two-loop topologies entering the 2-loop 6-point amplitudes. The arrow on the external line indicates leg number $1$.
\label{2loopints}}\nonumber
\end{figure}

\subsection{Supercut Example: the Double-Pentagon Cut}
%; towards the 6-point NMHV amplitude

Let us illustrate the general
strategy outlined in the previous section, by examining in some
detail two cuts that are essential for the construction of the
six-point NMHV superamplitude.  We begin with the `\dpc{}' cut, shown in
fig.~\ref{2loopcuts}(a), which isolates the double-pentagon
integrals $\Int{12}$ and $\Int{13}$ (shown in fig.~\ref{2loopints})
from a wide class of other integrals
(hence its name).

\renewcommand{\thesubfigure}{(\roman{subfigure})}
\begin{figure}
  \centering
  \subfigure[]{%
    \includegraphics[scale=0.5]{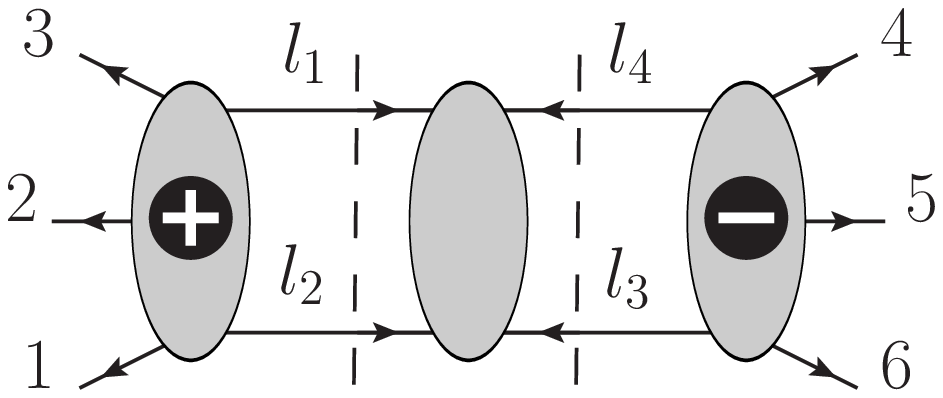}
    \label{cutacomponents_i}
  }
  \subfigure[]{%
    \includegraphics[scale=0.5]{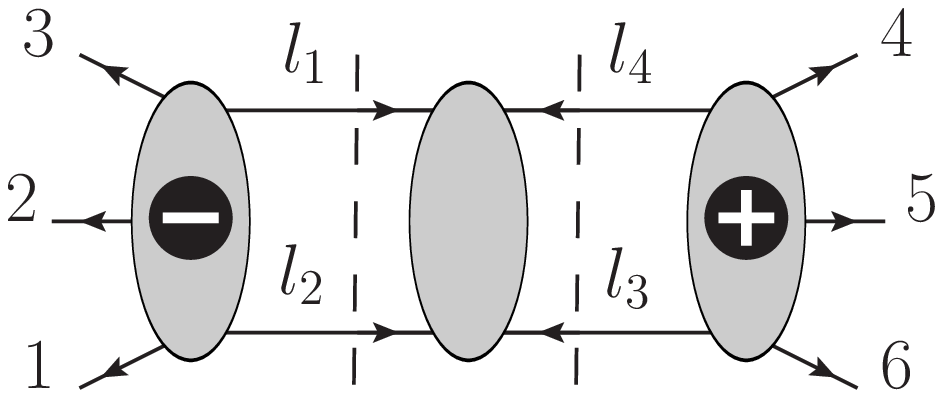}
    \label{cutacomponents_ii}
  }
  \caption{The two contributions to the `\dpc{}' supercut~\ref{2loopcuts}(a). The circled $+$ and $-$ denote MHV and
  ${\overline {\rm MHV}}$ superamplitudes, respectively. The middle
  amplitude may be chosen to be either of MHV or of ${\overline {\rm
      MHV}}$ type. Here we choose to present it as an MHV
  superamplitude.}
  \label{cutacomponents}
\end{figure}

\def\dps{{\rm dp}}
This cut provides two distinct contributions to the coefficients of
the NMHV amplitude, depending on which of the two five-point
tree-level factors is an MHV or an ${\overline {\rm MHV}}$
superamplitude.  Each of the contributions is closed under
supersymmetry transformations, so we will call them supersectors.
(They were called ``holomorphicity configurations'' in 
ref.~\cite{Bern:2009xq}.)
 These two supersectors are shown in fig.~\ref{cutacomponents};
their values are,
\be
&&{\cal C}^{\dps}_{\ref{cutacomponents_i}}=
\int d^4\eta_{l_1}d^4\eta_{l_2}
        d^4\eta_{l_4}d^4\eta_{l_3}d^8\omega\nonumber\\
&&\hskip 20mm\times
        \frac{\delt8(q_{123}^{A}+\lambda_{l_1}\eta^A_{l_1}
        +\lambda_{l_2}\eta^A_{l_2})}
        {\langle12\rangle\langle23\rangle\langle3 l_1\rangle
          \langle l_1 l_2\rangle \langle l_2 1\rangle}
          \frac{\delt8(\lambda_{l_1}\eta^A_{l_1}
        +\lambda_{l_2}\eta^A_{l_2}+\lambda_{l_4}\eta^A_{l_4}
        +\lambda_{l_3}\eta^A_{l_3})}
        {\langle l_2l_1\rangle\langle l_1 l_4\rangle\langle l_4 l_3\rangle\langle l_3 l_2\rangle}
        \\
&&\hskip 20mm\times
	\frac{\delt4(\eta_{l_4}^A-{\tilde\lambda}_{l_4}^{\dot\alpha}\omega^A_{\dot\alpha})
	\delt4(\eta_{l_3}^A-{\tilde\lambda}_{l_3}^{\dot\alpha}\omega^A_{\dot\alpha})}
	{[45][56][6l_3][l_3l_4][l_4 4]}
	\prod_{i=4}^6\delt4(\eta_i^A-{\tilde\lambda}_i^{\dot\alpha}\omega^A_{\dot\alpha})\nonumber
\ee
and
\be
&&{\cal C}^{\dps}_{\ref{cutacomponents_ii}}=
\int d^4\eta_{l_1}d^4\eta_{l_2}
        d^4\eta_{l_4}d^4\eta_{l_3}d^8\omega\nonumber\\
&&\hskip 20mm\times
        \frac{\delt8(q_{456}^{A}+\lambda_{l_4}\eta^A_{l_4}
        +\lambda_{l_3}\eta^A_{l_3})}
        {\langle45\rangle\langle56\rangle\langle6 l_3\rangle
          \langle l_3 l_4\rangle \langle l_4 4\rangle}
          \frac{\delt8(\lambda_{l_1}\eta^A_{l_1}
        +\lambda_{l_2}\eta^A_{l_2}+\lambda_{l_4}\eta^A_{l_4}
        +\lambda_{l_3}\eta^A_{l_3})}
        {\langle l_2l_1\rangle\langle l_1 l_4\rangle\langle l_4 l_3\rangle\langle l_3 l_2\rangle}
        \\
&&\hskip 20mm\times
	\frac{\delt4(\eta_{l_1}^A-{\tilde\lambda}_{l_1}^{\dot\alpha}\omega^A_{\dot\alpha})
	\delt4(\eta_{l_2}^A-{\tilde\lambda}_{l_2}^{\dot\alpha}\omega^A_{\dot\alpha})}
	{[12][23][3l_1][l_1l_2][l_2 1]}
	\prod_{i=1}^3\delt4(\eta_i^A-{\tilde\lambda}_i^{\dot\alpha}\omega^A_{\dot\alpha})\nonumber
\ee
where
\be
q_{123}^{\alpha A}=\sum_{i=1}^3\lambda_i^\alpha\eta^A_i
\quad\text{and}\quad
q_{456}^{\alpha A}=\sum_{i=4}^6\lambda^\alpha_i\eta^A_i\,.
\ee
By taking appropriate linear combinations of the arguments of the 
delta functions in these equations it is easy to extract the overall
super-momentum conservation delta function,
$\delt8(q_{123}^{A}+q_{456}^{A})$. 
Carrying out the Grassmann integrals we then find
%%%%%%%%%%%%%%%%%%%%%%%%%%%%%%%%%%%
\be
\label{caa}
{\cal C}^{\dps}_{\ref{cutacomponents_i}}&=&\sA^{(0),\text{MHV}} R_{146}\;
\frac{s_{123}\langle 1|2+3|4] \langle 3|4+5|6]
\langle l_1 l_2\rangle^4}
        {(\langle3 l_1\rangle
          \langle l_1 l_2\rangle \langle l_2 1\rangle)\,
          (\langle l_2l_1\rangle\langle l_1 l_4\rangle\langle l_4 l_3\rangle\langle l_3 l_2\rangle)\;
          ([6l_3][l_3 l_4][l_4 4])}\,,
%\nonumber
\\
\label{cab}
{\cal C}^{\dps}_{\ref{cutacomponents_ii}}&=&
\sA^{(0),\text{MHV}} R_{413}
\frac{s_{123}\langle 4|2+3|1] \langle 6|4+5|3] \, \langle l_4 l_3\rangle^4}
        {(\langle6 l_3\rangle
          \langle l_3 l_4\rangle \langle l_4 4\rangle)\;
          (\langle l_2l_1\rangle\langle l_1 l_4\rangle\langle l_4 l_3\rangle\langle l_3 l_2\rangle)\;
          ([3l_1][l_1l_2][l_2 1])}\,.
\ee
We can reorganize the contributions into even and odd components,
\be
{\cal C}^{\dps}={\cal C}^{\dps}_{\ref{cutacomponents_i}}+ {\cal C}^{\dps}_{\ref{cutacomponents_ii}}
    = \sA^{(0),\text{MHV}}\; {\cal C}^{\dps}_+\; (R_{413} + R_{146} )
     + \sA^{(0),\text{MHV}} \; {\cal C}^{\dps}_- \; (R_{413} - R_{146} )\,.
 \ee
 
The two functions of vanishing weight in equations (\ref{caa}) and
(\ref{cab}) may be identified as the contribution of gluon
intermediate states in a component approach.
They can be decomposed by standard means, by reconstructing
propagators and organizing the numerator into a single
trace\footnote{This last step is important to avoid the appearance of
  parity-even terms which are a product of two parity-odd factors.}\hskip -5pt.
For example, 
%%%%%%%%%%%%%%%%%%%%%%%%%%%%
\be
{\cal C}^{\dps}_+ &=&
\frac{s_{123}}
        {(\langle l_2l_1\rangle\langle l_1 l_4\rangle\langle l_4 l_3\rangle\langle l_3 l_2\rangle)} \biggl(
\frac{\langle 1|2+3|4] \langle 3|4+5|6]
\langle l_1 l_2\rangle^4}
        {(\langle3 l_1\rangle
          \langle l_1 l_2\rangle \langle l_2 1\rangle)\,
          ([6l_3][l_3 l_4][l_4 4])}
\nonumber\\
&&\hskip 42mm
+ \frac{\langle 4|2+3|1] \langle 6|4+5|3] \, \langle l_4 l_3\rangle^4}
        {(\langle6 l_3\rangle
          \langle l_3 l_4\rangle \langle l_4 4\rangle)\;
          ([3l_1][l_1l_2][l_2 1])}
\biggr)\nonumber\\
%&&
%&&
%\frac{s_{123}\langle 4|2+3|1] \langle 6|4+5|3] \, \langle l_4 l_3\rangle^4}
%        {(\langle6 l_3\rangle
%          \langle l_3 l_4\rangle \langle l_4 4\rangle)\;
%          (\langle l_2l_1\rangle\langle l_1 l_4\rangle\langle l_4 l_3\rangle\langle l_3 l_2\rangle)\;
%          ([3l_1][l_1l_2][l_2 1])}
%
%\frac{\Tr_+[k_1l_2l_1k_3k_{123}k_6 l_3 l_4 k_4 k_{456}]}
%{(2l_2\cdot k_1)(2l_1\cdot k_3)(2l_2\cdot l_3)(2l_4\cdot k_4)(2l_3\cdot k_6)}=
%
&=&\frac{1}{4}\biggl[
\frac{s_{123} ( s_{123} s_{345}-s_{12} s_{45})}
{(l_1+k_3)^2(l_2+l_3)^2  (l_3+k_6)^2}
+\frac{s_{61} s_{123}^2 }
{(l_2+k_1)^2 (l_2+l_3)^2 (l_3+k_6)^2}
\cr
&&
+\frac{s_{34} s_{123}^2}
{(l_1+k_3)^2(l_2+l_3)^2 (l_4+k_4)^2 }
+\frac{s_{123}( s_{123} s_{234}-s_{23} s_{56})}
{(l_2+k_1)^2 (l_2+l_3)^2 (l_4+k_4)^2 }
\cr
&&
+\frac{s_{12} s_{23} s_{123} (k_6- l_2)^2}
{(l_2+k_1)^2 (l_1+k_3)^2(l_2+l_3)^2  (l_3+k_6)^2}
+\frac{s_{12} s_{23} s_{123} (k_4- l_1)^2}
{(l_2+k_1)^2 (l_1+k_3)^2(l_2+l_3)^2 (l_4+k_4)^2 }
\cr
&&
+\frac{s_{45} s_{56} s_{123} (k_3- l_4)^2}
{(l_1+k_3)^2(l_2+l_3)^2 (l_4+k_4)^2 (l_3+k_6)^2}
+\frac{s_{45} s_{56} s_{123} (k_1- l_3)^2}
{(l_2+k_1)^2 (l_2+l_3)^2 (l_4+k_4)^2 (l_3+k_6)^2}
\cr
&&
+\frac{1}{(l_2+k_1)^2 (l_1+k_3)^2(l_2+l_3)^2 (l_4+k_4)^2 (l_3+k_6)^2}
\cr
&&\qquad
\times
 s_{123}\Big(
(s_{12} s_{45} -  s_{123} s_{345}) (k_1-l_3)^2(k_4- l_1)^2
+s_{34} s_{123} (k_1- l_3)^2(k_6- l_2)^2
\cr
&&\qquad\qquad
+s_{61} s_{123} (k_3- l_4)^2(k_4- l_1)^2
+(s_{23} s_{56}  - s_{123} s_{234}) (k_3- l_4)^2(k_6- l_2)^2
\Big)
\cr
&&
+\frac{1}{(l_2+k_1)^2 (l_1+k_3)^2 (l_4+k_4)^2 (l_3+k_6)^2}
\\
&&\qquad
\times
\Big( 2s_{12} s_{23} s_{45} s_{56}- s_{123} \big(s_{61} s_{34} s_{123}
    + s_{12} s_{45} s_{234} +
       s_{23} s_{56} s_{345} -
       s_{123} s_{234} s_{345}  \big)
       \Big)
\biggr]
\nonumber
\ee
From this expression we can easily read off the coefficients of all
integrals in fig.~\ref{2loopints} that have a double cut in the
$s_{123}$ channel. Some integrals appear multiple times, corresponding
to different cyclic permutations of external legs that have such a
cut.
The numerator factors in the expression above are precisely those
required to render the integrals invariant under dual inversion. As we
did not need to specify the helicity labels of the external legs, all
cuts with this topology can be obtained by simple cyclic relabeling.

The `turtle' cut shown in fig.~\ref{2loopcuts}({\it b\/}) 
can be computed in a similar
way, and also contributes a lone $R$ invariant. These two cuts
determine the coefficients of all integral topologies in
fig.~\ref{2loopints} except $\Int7$, $\Int{14}$, and $\Int{15}$.
Other cuts are necessary to determine these contributions.
An efficient strategy, which makes use of the results obtained from the 
\dpc{} cut of fig.~\ref{2loopcuts}({\it a\/}) and 
the turtle cut of fig.~\ref{2loopcuts}({\it b\/}), 
is to analyze the relevant next-to-maximal 
cuts and find the remaining integrals one at a time.

\begin{figure}
  \centering
  \subfigure[]{%
    \includegraphics{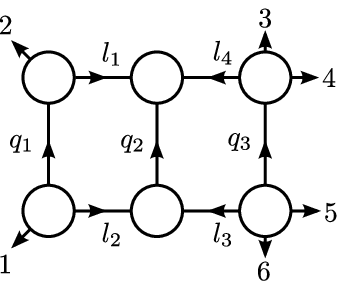}
    \label{fig:oott_template}
  }
  \subfigure[]{%
    \includegraphics{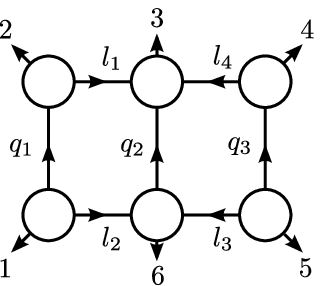}
    \label{fig:turtle_template}
  }
  \subfigure[]{%
    \includegraphics{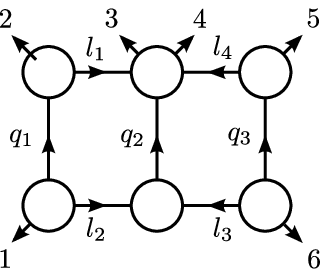}
    \label{fig:ootoo_template}
  }
  \caption{Next-to-maximal cuts that detect the integrals not 
easily isolated by the iterated two-particle cuts: from left to right,
next-to-maximal cuts for the `hexabox,' `\sqrc,' and `rabbit-ears' cuts
of fig.~\ref{2loopcuts}(c), (d), and (e), respectively.}
  \label{fig:hard-cuts}
\end{figure}

\begin{figure}
  \centering
  \subfigure[]{%
    \includegraphics{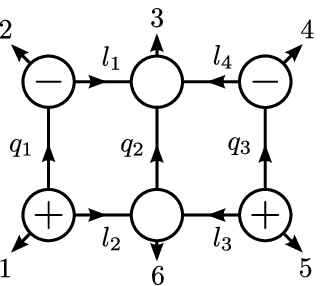}
    \label{fig:turtle1}
  }
  \subfigure[]{%
    \includegraphics{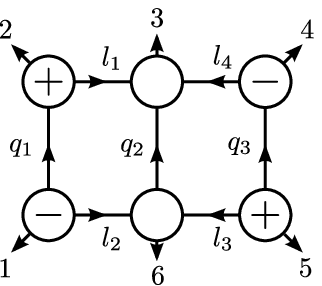}
    \label{fig:turtle2}
  }
  \subfigure[]{%
    \includegraphics{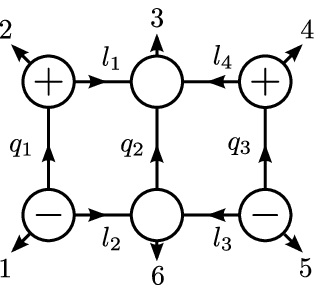}
    \label{fig:turtle3}
  }
  \subfigure[]{%
    \includegraphics{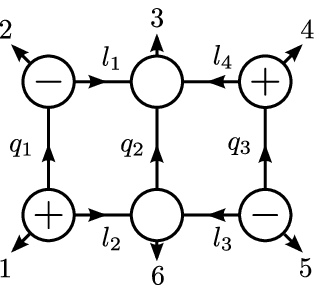}
    \label{fig:turtle4}
  }
  \caption{The four possible assignments of internal helicities for
    the next-to-maximal
    `\sqrc' cut of fig.~\ref{fig:turtle_template}.  The `$\ominus$'
    vertices denote three-point $\overline{\text{MHV}}$ amplitudes
    while the `$\oplus$' vertices denote three-point MHV amplitudes.}
  \label{fig:turtle_cases}
\end{figure}

%%%%%
\subsection{Supercut Example: A Contribution to the Flying-Squirrel Cut}
%\subsubsection{A component of cut in fig. \ref{2loopcuts}(d)}

Following this strategy, we present one contribution to the
next-to-maximal cut, shown in fig.~\ref{fig:turtle_template}, that
imposes additional cut constraints beyond the `\sqrc' cut of
fig.~\ref{2loopcuts}({\it d\/}).  It serves to isolate one of our target
integrals, $\Int7$, and allows us to determine its coefficient.  As
explained above, we impose the additional cut conditions because of
difficulties in organizing the results for cuts like the `\sqrc' cut in
terms of dual superconformal $R$ invariants.
A superspace calculation, combined with the reduced set of
Feynman integrals isolated by
the next-to-maximal cut conditions,
reduces the ambiguity in such a reorganization by enforcing
supersymmetry relations between the reduced number of contributions to
these cuts.

The next-to-maximal `\sqrc' cut of fig.~\ref{fig:turtle_template} has four
supersectors, corresponding to three-point amplitudes at the corners
being of MHV or \MHVbar{} type.  The supersectors are shown in
fig.~\ref{fig:turtle_cases}.  We discuss in detail the configuration
in fig.~\ref{fig:turtle2}, quote the result for the configuration in
fig.~\ref{fig:turtle1}, and explain how to construct the other two
components by relabeling.

The product of the tree superamplitudes entering the supersector shown in  
fig.~\ref{fig:turtle2} is
\begin{multline}
 {\cal C}_{\ref{fig:turtle2}}= \int d^4 \eta_{l_1} d^4 \eta_{l_2} d^4 \eta_{l_3} d^4 \eta_{l_4} d^4 \eta_{q_1} d^4 \eta_{q_2}
  d^4 \eta_{q_3} 
 \frac{\delt4([1 q_1] \eta^A_{l_2} + [q_1 l_2] \eta^A_1 + [l_2 1] \eta^A_{q_1})}
  {[1 q_1] [q_1 l_2] [l_2 1]} \\
  \times
  \frac{\delt8(-\lambda_{q_1} \eta^A_{q_1} + \lambda_2 \eta^A_2 + \lambda_{l_1} \eta^A_{l_1})}
  {\langle q_1 2\rangle \langle 2 l_1\rangle \langle l_1 q_1\rangle} 
  \frac{\delt8(-\lambda_{l_1}\eta^A_{l_1} + \lambda_3 \eta^A_3 - \lambda_{l_4} \eta^A_{l_4} 
  - \lambda_{q_2} \eta^A_{q_2})}{\langle l_1 3\rangle \langle 3 l_4\rangle \langle l_4 q_2\rangle 
  \langle q_2 l_1\rangle} \label{comp2d}\\
 \times
  \frac{\delt4([l_4 4] \eta^A_{q_3} + [4 q_3] \eta^A_{l_4} + [q_3 l_4] \eta^A_4)}{[l_4 4] [4 q_3] [q_3 l_4]}
  \frac{\delt8(\lambda_{q_3} \eta^A_{q_3} + \lambda_5 \eta^A_5 + \lambda_{l_3} \eta^A_{l_3})}
  {\langle q_3 5\rangle \langle 5 l_3\rangle \langle l_3 q_3\rangle} \\
  \times
  \frac{\delt8(\lambda_6 \eta^A_6 - \lambda_{l_2} \eta^A_{l_2} + \lambda_{q_2} \eta^A_{q_2}
  - \lambda_{l_3} \eta^A_{l_3})}{\langle 6 l_2\rangle \langle l_2 q_2\rangle \langle q_2 l_3\rangle
  \langle l_3 6\rangle}.
\end{multline}
The expected overall 
super-momentum conservation constraint may be extracted by adding the
arguments of all the eightfold delta functions $\delt8$ to the last such 
function, and then using momentum conservation to eliminate $\lambda_{q_1}
\eta^A_{q_1}$ and $\lambda_{l_4} \eta^A_{l_4}$.
These transformations have unit Jacobian. 

The remaining Grassmann integrals can be computed easily; we obtain:
\be
{\cal N}\!\!&=&\!
[1 q_1]^4 \delt4\big(
\langle q_1 2\rangle \langle q_2 l_1\rangle\langle q_3 l_3\rangle  [4 q_3]\eta_2^A
+
\langle q_1 l_1\rangle \langle q_2 3\rangle \langle q_3 l_3\rangle  [4 q_3]\eta_3^A
\nonumber\\[1pt]
&&\qquad\quad\;
+
\langle q_1 l_1\rangle \langle q_2 l_4\rangle\langle q_3 l_3\rangle  [q_3 l_4]\eta_4^A
+
\langle q_1 l_1\rangle \langle q_2 l_4\rangle\langle l_3 5\rangle  [l_4 4]\eta_5^A
\big)~~.
\label{numerator}
\ee
The contribution from the supersector in fig.~\ref{fig:turtle2}
 is obtained by dividing ${\cal N}$ 
by the explicit denominators in equation~(\ref{comp2d}). 
This expression can be simplified in several different ways; 
we proceed by solving the cut 
conditions. 
The internal spinors (except for those associated to $q_2$)
may be expressed conveniently in terms of two variables $y$ and $z$:
\be
\label{kinematicsd}
\begin{tabular}{lllll}
 			    $\lambda_{l_1} = y \lambda_1 - \lambda_2\,,$
&~~~~~~~~~~~~&$\tlambda_{l_1} = \tlambda_2\,,$
&~~~~~~~~~~~~&$\lambda_{q_1} = y \lambda_1\,,$
\\
			    $ \lambda_{l_2} = -\lambda_1\,,$
&~~~~~~~~~~~~&$\tlambda_{l_2} = \tlambda_1 + y \tlambda_2\,,$
&~~~~~~~~~~~~&$\tlambda_{q_1} = \tlambda_2\,,$
\\
			    $\lambda_{l_3} = -z \lambda_4 - \lambda_5\,,$
&~~~~~~~~~~~~&$\tlambda_{l_3} = \tlambda_5\,,$
&~~~~~~~~~~~~&$\lambda_{q_3} = z \lambda_4\,,$
\\
			    $\lambda_{l_4} = \lambda_4\,,$
&~~~~~~~~~~~~&$\tlambda_{l_4} =   -\tlambda_4+ z \tlambda_5\,,$
&~~~~~~~~~~~~&$\tlambda_{q_3} = \tlambda_5\,$.
\end{tabular}
\ee
The momentum $q_2$ can be determined through momentum conservation;  the condition that
it be on shell relates the two parameters $y$ and $z$.
These relations imply that all spinor products in
eq.~(\ref{numerator}) that do not contain the holomorphic spinor
$|q_2\rangle$ are monomials in $y$, $z$ and spinor products of
external momenta.
%
% and in fact are common to all the terms in the argument of the delta 
%function. 
%
The remaining holomorphic spinor products, which do contain
$|q_2\rangle$, can be converted into functions of $y$ and external
spinor products by multiplying and dividing by $[q_2 5]^4$ and using
the identities,
\begin{eqnarray}
\label{q2identities}
  \langle l_1 q_2\rangle [q_2 5] &=& - \langle 2(3+4)5] + y \langle   1(3+4)5]\,,\cr
  \langle l_2 q_2\rangle [q_2 5] &=& \langle 1 6\rangle [6 5]\,,
  \\
  \langle l_3 q_2\rangle [q_2 5] &=& \langle 1 6\rangle ([1 6] + y [2   6])\,,\cr
  \langle l_4 q_2\rangle [q_2 5] &=& \langle 4(2+3)5] + y \langle 1 4\rangle [2 5]\,.
  \nonumber
\end{eqnarray}
The numerator factor ${\cal N}$ 
in eq.~\eqref{numerator} is then,
\begin{multline}
{\cal N}= \frac{s_{12}^4 s_{45}^4 y^4 z^4}{[q_2 5]^4}
  \delt4\Bigl(\eta^A_2 (\langle 2(1+6)5] - y \langle 1(2+6)5])
  + \eta^A_3 (\langle 3(1+6)5] - y \langle 1 3\rangle [2 5]) \\
  + \eta^A_4 (\langle 4(1+6)5] - y \langle 1 4\rangle [2 5])
  + \eta^A_5 (s_{234} - y \langle 1(3+4)2])\Bigr).
\label{leftoverdelta}
\end{multline}
This expression is invariant, though not manifestly, under the
action of the supersymmetry generators $\bar{q} = \sum
\tlambda \partial_\eta$.

Overall super-momentum conservation provides the means to further simplify ${\cal N}$.
By subtracting $\sum_i \eta^A_i \langle i(1+6)5] = 0$, and adding  
$\sum_i y \eta^A_i \langle 1 i\rangle [2 5] = 0$ to the argument of the 
delta function we find
\begin{equation}
 \delt8(\sum_i\lambda_i\eta_i) {\cal N}=  
 s_{12}^4 s_{45}^4 y^4 z^4\frac{\langle 1 6\rangle^4}{[q_2 5]^4} 
 \delt4(\eta^A_1 [56] + \eta^A_5 [61] + \eta^A_6 [15]
  + y (\eta^A_2 [56] + \eta^A_5 [62] + \eta^A_6 [25]))~~.
\label{leftoverdelta_mod}
\end{equation}  

This superspace expression has the two unwanted features already
mentioned in section~\ref{unitarity_general}: on the one hand, the
argument of the delta function depends on the internal momenta through
the variable $y$; on the other, it is not manifestly a function only
of the same superspace structures as the tree-level amplitude
(\ref{NMHVampAllLoop}). We wish to reorganize it in terms of $R$ invariants,
and at the same time remove the dependence on internal momenta from
the arguments of the delta function by using the Lagrange
interpolation formula (\ref{LIf}) on the degree four polynomial,
\be
P_4(y)&=&\delt4(\eta^A_1 [56] + \eta^A_5 [61] + \eta^A_6 [15]
  + y (\eta^A_2 [56] + \eta^A_5 [62] + \eta^A_6 [25]))
  \cr
  &=&
  \sum_{i=1}^5 \prod_{\substack{j=1\\j \neq i}}^5 \frac{y-y_j}{y_i - y_j} P_4(y_i)\,.
\ee
For this to be possible, as explained earlier
it is necessary that there exist at least five values $y_i$ such that
$P_4(y_i)$ is proportional to an $R$ invariant. It turns out that there are at
least six such values:
\begin{subequations}
\label{eq:delta4values}
\begin{align}
  P_4\left(\frac{\langle 2(3+4)5]}{\langle 1(3+4)5]}\right) &=
  \left(\frac{\langle 3 4\rangle [5 6] }{\langle 1(3+4)5]}\right)^4
  \delt4(\eta^A_3 [4 5] + \eta^A_4 [5 3] + \eta^A_5 [3 4])
  \propto R_{635}\,,\\
  P_4\left(-\frac{[1 6]}{[2 6]}\right) &= \left(\frac{[5 6]}{[2 6]}\right)^4
  \delt4(\eta^A_6 [1 2] + \eta^A_1 [2 6] + \eta^A_2 [6 1])
  \propto R_{362}\,,\\
  P_4\left(-\frac{\langle 4(5+6)1]}{\langle 4(5+6)2]}\right) &=
  \left(\frac{\langle 3 4\rangle [5 6]}{\langle 4(5+6)2]}\right)^4
  \delt4(\eta^A_1 [2 3] + \eta^A_2 [3 1] + \eta^A_3 [1 2])
  \propto R_{413}\,,\\
  P_4\left(\frac{\langle 2 3\rangle}{\langle 1 3\rangle}\right) &=
  \left(\frac{\langle 3 4\rangle}{\langle 1 3\rangle}\right)^4
  \delt4(\eta^A_4 [5 6] + \eta^A_5 [6 4] + \eta^A_6 [4 5])
  \propto R_{146}\,,\\
  P_4\left(\frac{s_{234}}{\langle 1(3+4)2]}\right) &=
  \left(\frac{\langle 3 4\rangle [5 6]}{\langle 1(3+4)2]}\right)^4
  \delt4(\eta^A_2 [3 4] + \eta^A_3 [4 2] + \eta^A_4 [2 3])
  \propto R_{524}\,,\\
  P_4(0) &=  \delt4(\eta^A_5 [6 1] + \eta^A_6 [1 5] + \eta^A_1 [5 6])
  \propto R_{251}\,.
\end{align}
\end{subequations}
For some of these cases, we have used
overall super-momentum conservation constraint 
as well as nontrivial spinor identities
to transform the argument of $\delt4$. 
As we will see shortly, only the first four values of $y$ correspond to leading 
singularities.

We can use the Lagrange interpolation formula for any five of the six special values 
\be
\label{eq:six_values}
\{y_1,y_2,y_3,y_4,y_5,y_6\}=
\biggl\{\frac{\langle   2(3+4)5]}{\langle 1(3+4)5]}, -\frac{[1 6]}{[2 6]}, 
-\frac{\langle 4(5+6)1]}{\langle 4(5+6)2]}, \frac {\langle 2 3\rangle}{\langle 1 3\rangle}, 
\frac{s_{234}}{\langle 1(3+4)2]}, 0\biggr\}~~.
\ee
Clearly, the decomposition obtained this way is not unique as there are six 
different possibilities. Let us denote them by $\mathcal{L}_i$, 
where $i$ is the index of the missing root.  In general we can construct a 
five-parameter decomposition 
\begin{equation}
  P_4(y) = \sum_{i=1}^6 \alpha_i \mathcal{L}_i(y)\,,
\end{equation}
with $\sum \alpha_i = 1$. The remaining parameters $\alpha_i$ may be
constrained by requiring that the superamplitude have additional
manifest symmetries; for example, one may require that the parity of
the superamplitude be manifest. We impose such a requirement in
the following.

All the presentations of the cut obtained for different possible
choices of five values $y_i$ are physically equivalent. However, they
contain different $R$ invariants; the
existence of more than five values $y_i$ is equivalent to the
existence of nontrivial relations between $R$ invariants. These
relations hold only in the presence of the overall super-momentum
conservation constraint.

We are now in position to assemble the result ${\cal
  C}_{\ref{fig:turtle2}}$ for the supersector of
fig.~\ref{fig:turtle2}.  Further use of the identities
(\ref{q2identities}) implies that it is given by
\begin{equation}
  \frac{-\sA_{6}^{(0),\text{MHV}}\,s_{12} s_{45}\;P_4(y)}{\langle 3 4\rangle [5 6] 
  (y \langle 1 3\rangle - \langle 2 3\rangle)(y \langle 1(3+4)5] - \langle 2(3+4)5]) 
  (\langle 4(5+6)1] + y \langle 4(5+6)2]) ([1 6] + y [2 6])}~~.
\end{equation}
Inspecting the denominator of this expression, we see that the first four
points $y_i$ in eq.~(\ref{eq:six_values}) correspond to poles.
 They are in fact positions of leading singularities,
as all of them arise from the $[q_2 5]^{-4}$ factor in
eq.~(\ref{leftoverdelta_mod}) which is the Jacobian arising from
solving the cut conditions.

Choosing the five points $y_i$ to be $\lbrace y_1, y_2, y_3, y_4, y_5\rbrace$, 
we obtain,
\begin{multline}
  \label{eq:hard-cut}
{\cal C}_{\ref{fig:turtle2}}=  s_{12} s_{45} s_{234} \sA_{6}^{(0),\text{MHV}}\,
\left(\;\frac {y_4 (y-y_5)}{y_5 (y-y_4)} R_{146} 
         + \frac {y_2 (y-y_5)}{y_5 (y-y_2)} R_{362} \right.
\\
\left.- \frac {y_3 (y-y_5)}{y_5 (y-y_3)} R_{413} 
         - \frac {y_1 (y-y_5)}{y_5 (y-y_1)} R_{635} + R_{524}
\right)\,.
\end{multline}
Each of the denominators appearing in this expression may be identified with a propagator
evaluated on the kinematic configuration~(\ref{kinematicsd}).

Thus, the contribution of this supersector depends only on
$R$ invariants.
We can decompose it in 
even and odd invariants $(R_{i,i+3,i+5}\pm
R_{i+3,i,i+2})$, following the
form~(\ref{NMHVampAllLoop}) of the superamplitude.
To identify the part of ${\cal C}_{\ref{fig:turtle2}}$ that receives
contributions from the missing integral $\Int7$ we need to subtract from
it the contribution of all the other integrals in
fig.~\ref{2loopints}, determined from the cuts
of fig.~\ref{2loopcuts}(a) and~(b).  
These cuts can contribute only terms 
proportional to the invariants $R_{146}$, $R_{362}$, $R_{413}$ or
$R_{635}$. Thus, we can conclude immediately that $R_{524}$ 
arises solely as a coefficient of $\Int7$, whose coefficient must
therefore be,
\begin{equation}
  \frac{1}{2} 
\sA_6^{(0),\text{MHV}}\,s_{12} s_{45} s_{234} (R_{251} + R_{524})\,.
\end{equation}  
Indeed, it is intuitively clear that because of its topology, 
 $\Int7$ can appear in the coefficient~\eqref{eq:hard-cut}
only in terms that have no additional propagators.
 
Carefully repeating this analysis for the other even invariants
implies that the complete contribution of this supersector
to the even part of $\Int7$'s coefficient is,
\be
  \label{eq:turtle_coeff}
  &&-\frac 1 4 \sA_6^{(0),\text{MHV}}(R_{146} + R_{413}) s_{123} (s_{234} s_{345} - s_{61} s_{34})\cr
  &&- \frac 1 4 \sA_6^{(0),\text{MHV}}(R_{362} + R_{635}) s_{345} (s_{123} s_{234} - s_{23} s_{56}) \cr
  &&+ \frac 1 2 \sA_6^{(0),\text{MHV}}(R_{251} + R_{524}) s_{12} s_{45} s_{234}\,.
\ee
This conclusion must be checked against the other configurations in
fig.~\ref{fig:turtle_cases}.  Fig.~\ref{fig:turtle4} is the parity
conjugate of fig.~\ref{fig:turtle2} and should therefore yield the
same result for the even $R$ invariants (and its negative
for the  odd-parity ones).

The configurations in figs.~\ref{fig:turtle1} and~\ref{fig:turtle3}
and are parity conjugates of each other. Evaluating them following the
same steps yields,
\be
\label{turtle_case1}
 {\cal C}_{\ref{fig:turtle1}}= 
 \sA_6^{(0),\text{MHV}} \frac{s_{12} s_{45} \delt4([6 1] \eta^A_5 + [1 5] \eta^A_6 + [5 6] \eta^A_1)}
  {\langle 2 3\rangle \langle 3 4\rangle [5 6] [6 1] \langle 4(2+3)1] \langle 2(1+6)5]}
  &=& \sA_6^{(0),\text{MHV}} R_{251} s_{12} s_{45} s_{234}\,,~~~
  \\[2pt]
 {\cal C}_{\ref{fig:turtle3}}=
   \sA_6^{(0),\text{MHV}} \frac{s_{12} s_{45} \delt4([3 4] \eta^A_2 + [4 2] \eta^A_3 + [2 3] \eta^A_4)}
  {\langle 1 6\rangle \langle 6 5\rangle [2 3] [3 4] \langle 5(1+6)2] \langle 1(2+3)4]}
  &=& \sA_6^{(0),\text{MHV}} R_{524} s_{12} s_{45} s_{234}\,.~~~
\label{turtle_case3}
\ee
Unlike the configuration in
fig.~\ref{fig:turtle2}, the loop-momentum dependence here cancels
completely after integration over the internal Grassmann variables.
One may verify that evaluating
\eqref{eq:turtle_coeff} on the relevant internal kinematic
configuration reproduces
eqs.~\eqref{turtle_case1}
and~\eqref{turtle_case3}.

The component of the \sqrc{} cut of fig.~\ref{2loopcuts}(d) that we
evaluated shows that this cut contributes to all two-loop scalar
functions $W_i^{(2)}$.  The same is true for the other cuts with this
topology but with cyclicly-permuted external legs.  The sum over
cyclic permutations may be reorganized in terms of permutations of a
single ``even'' spin coefficient $R_{146} + R_{413}$ which multiplies
three different integrals of the type $\Int7$ with different
assignments of external legs. We will use this presentation in the
following section.

\section{The Two-Loop Six-Point Superamplitude%
 \label{2loopintegrands}}

\begin{figure}[ht]
\centerline{\includegraphics[scale=0.45]{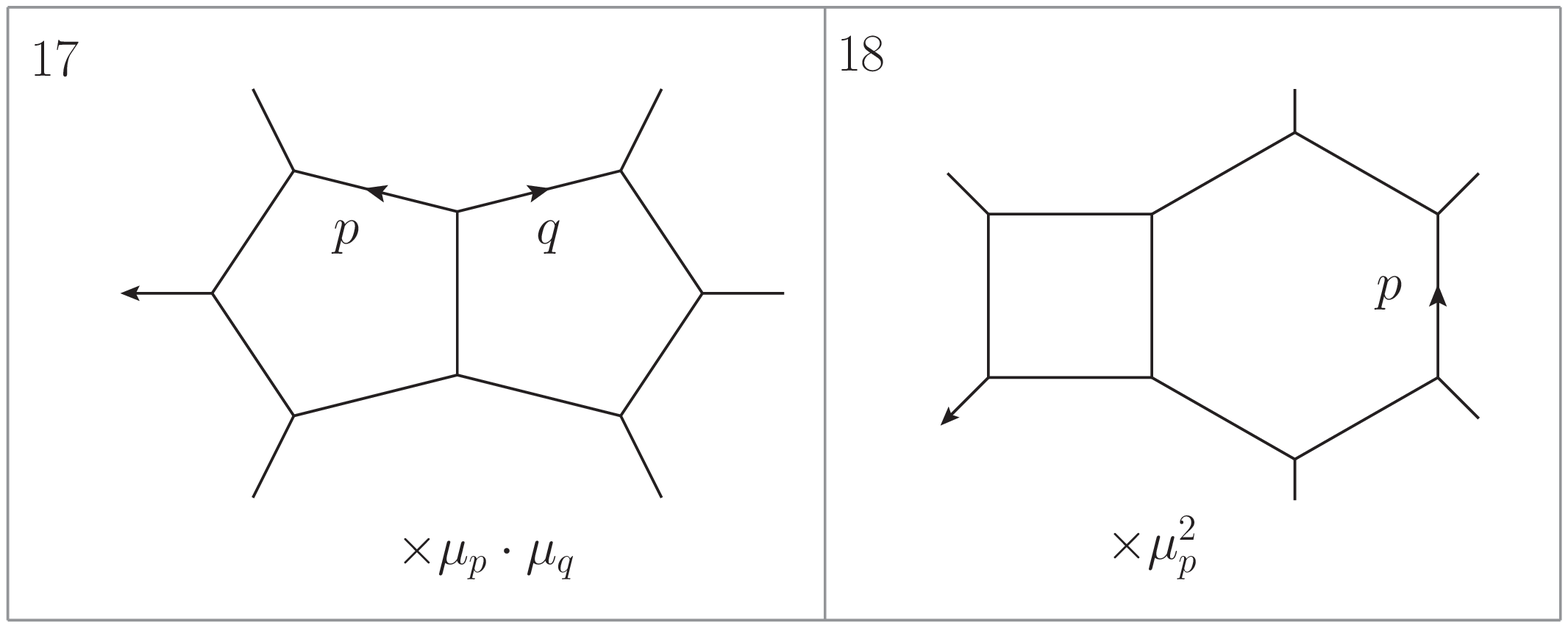}}
\caption{$\mu$-integrals entering the 2-loop 6-point amplitudes.
The arrow on the external line indicates leg number $1$.
\label{muints}}\nonumber
\end{figure}

We determined the four-dimensional cut-constructible even parts
$W_i^{(2),D=4}$ of the two-loop six-point NMHV superamplitude,
\be
W_i^{(2),\text{NMHV}}=W_i^{(2),D=4}+W_i^{(2),\mu},
\ee
as explained in the previous section,
by analyzing the cuts shown in fig.~\ref{2loopcuts}
or next-to-maximal versions of them.
We obtained an explicit expression for the remaining part, $W_i^{(2),\mu}$,
cut-constructible only in 
$D$-dimensions, by comparing
the results of $D$-dimensional and four-dimensional cut
calculations.
We have carried out the calculation without assuming a specific 
(possibly
overcomplete) basis of two-loop integrals and found that the integrals
listed in figs.~\ref{2loopints} and~\ref{muints} are necessary and
sufficient to saturate the cuts of the even part of the amplitude through
${\cal O}(\epsilon^0)$.
For comparison, and because of changes in the labeling of these integrals
with respect to the original calculations of the
two-loop six-point MHV amplitude~\cite{Bern:2008ap},
we also present it in our labeling.

\subsection{The NMHV amplitude}

The four-dimensional cut-constructible even part
of all six-point two-loop amplitudes is built out of
a sum of the 16 integrals
shown in fig.~\ref{2loopints}
\be
S^{(2),D=4}(\id)&=&
%(*1*)
\frac{1}{4}c_1 I^{(1)}(\epsilon)
%(*2*)
+c_2 I^{(2)}(\epsilon)
%(*3*)
+\frac{1}{2}c_3 I^{(3)}(\epsilon)
%(*4*)
+\frac{1}{2} c_{4} I^{(4)}(\epsilon)
%(*5*)
+c_{5} I^{(5)}(\epsilon)
%(*6*)
+c_{6} I^{(6)}(\epsilon)
\nonumber
\\[.5pt]
&+&
%(*7*)
%%+\frac{1}{2}
\frac{1}{4} \left(
%(*7a*)
   c_{7a}    \IP^{-2} I^{(7)}(\epsilon)
%(*7b*)
+ c_{7b} \IP^{-1} I^{(7)}(\epsilon)
%(*7c*)
+ c_{7c}  I^{(7)}(\epsilon)\right)
%(*8*)
+\frac{1}{2} c_{8} I^{(8)}(\epsilon)
%(*9*)
+c_{9} I^{(9)}(\epsilon)
\nonumber
\\[.5pt]
&+&
%(*10*)
c_{10} I^{(10)}(\epsilon)
%(*11*)
+c_{11} I^{(11)}(\epsilon)
%(*12*)
+\frac{1}{2} c_{12} I^{(12)}(\epsilon)
%(*13*)
+\frac{1}{2} c_{13} I^{(13)}(\epsilon)
\nonumber
\\[.5pt]
&+&
%(*16*)
\frac{1}{2} c_{14} I^{(14)}(\epsilon)
%(*17a*)
+\frac{1}{2} c_{15} I^{(15)}(\epsilon)
%(*18*)
+c_{16} I^{16}(\epsilon)\,.
%\nonumber
\label{Sblock}
\ee
The coefficients $c_i$, which differ between the MHV and the NMHV
amplitudes, are functions of external momenta and the numerical
coefficients are symmetry factors reflecting the symmetries of each
integral under cyclic permutations of external legs.

The functions $W_i^{(2),D=4}$ are constructed by summing $S^{(2),D=4}$
over the sets of permutations ${\cal S}_i$ in eqs.~\eqref{eq:perm1},
\eqref{eq:perm2} and~\eqref{eq:perm3} that map each
superinvariant $(R_{i+3,i,i+2}+R_{i,i+i,i+5})$ into itself:
\be
W_i^{(2),D=4}=\frac{1}{8}\sum_{\sigma\in {\cal S}_i}  S^{(2),D=4}(\sigma)+{\cal O}(\epsilon)\,.
\ee
Of the overall factor of $1/8$, a factor of $1/4$ emerges from the
calculation of the unitarity cuts and a factor of 1/2 is due to our
choice of normalization.

The coefficients $c_j$ in the identity permutation entering $S^{(2),D=4}$ 
in eq.~\eqref{Sblock} are:
\be
\begin{array}{rlcrl}
c_1 =& - s_{1 2 3}^2 s_{3 4} s_{6 1} + s_{1 2 3}^2 s_{2 3 4} s_{3 4 5}  &~~~
                         											  & c_2 =& 2 s_{1 2}^2 s_{2 3} \cr
        & - s_{1 2 3} s_{2 3 4} s_{1 2} s_{4 5}  -s_{1 2 3} s_{3 4 5} s_{2 3} s_{5 6}
            + 2 s_{1 2} s_{2 3} s_{4 5} s_{5 6}  
                                                                                                             &&          &                               \cr
c_3 =& s_{1 2 3} (s_{1 2 3} s_{3 4 5} - s_{1 2} s_{4 5})
&& c_4 = & s_{1 2 3}^2 s_{3 4}  \cr
c_5 =& - s_{1 2} s_{1 2 3} s_{2 3 4}                                                       &&c_6 =& s_{6 1} s_{1 2} s_{1 2 3}\cr
c_{7a} =&
-s_{1 2 3}  (s_{3 4 5} s_{2 3 4}-
%(1-f_4)
s_{6 1} s_{3 4})
%
% cuts fix -ffc2=ffc4=1/2; moved this 1/2 in symmetry factor
%
%         (- ffc2) s_{3 4 5} s_{1 2 3} s_{2 3 4}  +
%         (- ffc4) s_{1 2 3} s_{6 1} s_{3 4}
% ffc2 and ffc4 cancel in permutation sum against c_{7c}
         %(*7 *)
&&
c_{7b} =& 2s_{1 2 3} s_{3 4} s_{6 1} \cr
c_{7c} =&
%            f_2
-s_{1 2 3} (s_{2 3 4} s_{3 4 5} -
%f_4
 s_{6 1} s_{3 4})
%
% cuts fix -ffc2=ffc4=1/2; moved this 1/2 in symmetry factor
%
%c_{7c} &=& ffc2 s_{2 3 4} s_{3 4 5} s_{1 2 3}  +
%         ffc4 s_{1 2 3} s_{6 1} s_{3 4}
% cancel in permutation sum against ffc2 and ffc4 in c_{7a}
%                  %(*7 *)
&& c_8 =& 0        \cr
c_9 =& s_{1 2 3} s_{4 5} s_{5 6}          && 
c_{10} =& s_{5 6} s_{1 2 3} s_{3 4 5}   \cr
c_{11} =& -s_{5 6} s_{6 1} s_{1 2 3}
% (*11*)
&&
c_{12} =& -s_{1 2 3}( s_{1 2 3} s_{3 4 5} - s_{1 2} s_{4 5})
% (*12*)
\cr
c_{13} =& s_{1 2 3}^2 s_{6 1}
% (*13*)
&&
c_{14} =& 2 s_{3 4}^2 s_{1 2 3}
%(*16*)
\cr
c_{15} =& 0
%(*17*)
&&
c_{16} =& 2 s_{1 2} s_{3 4} s_{1 2 3}
%(*18*)
\cr
c_{17} =& {\rm not ~necessary}
&&
c_{18} =& \frac{1}{6}s_{123}(2s_{34}s_{61}-s_{234}s_{345})
\end{array}
\label{IntegralCoefficientsNMHV}
\ee
%%%%%%%%%%%%%%%%%%%%%%%%%%%%%%%%%%%%%%%%

In dimensional regularization, the six-point two-loop (and quite likely all
higher-point higher-loop) amplitudes receive contributions from
integrals --- collectively referred to as ``$\mu$-integrals'' --- whose integrand vanishes 
identically when evaluated in four
dimensions. The integrals shown in fig.~\ref{muints} are of this
type, where $\mu_p$ and $\mu_q$ denote the $(-2\epsilon)$ components
of the loop momenta.  As noted in ref.~\cite{Bern:2008ap}, the integral
$I^{(17)}$ vanishes identically as $\epsilon\rightarrow 0$; we will
therefore ignore it in the following.
To determine the contributions of such integrals we compare the result
of the four-dimensional cut calculation with that of the
$D$-dimensional cuts and find that the even part of the amplitude also
contains the terms,
\be W^{(2),\mu}_i&=&\Biggl(\sum_{\sigma\in{\cal
    S}_i} \frac{1}{4} c_{18}\Biggr) \sum_{\sigma\in {\cal S}_1\cup
  {\cal S}_2\cup {\cal S}_3}\,\frac{1}{2}s_{12}I^{(18)}(\sigma)\,.
\label{mu_ints_W}
\ee
The coefficients $c_j$ bear certain similarities to the
corresponding coefficients in the MHV amplitude.

\subsection{The MHV amplitude}

For completeness, and because of differences of notation from
ref.~\cite{Bern:2008ap}, we also present the integrand of the even part of
the MHV amplitude. The four-dimensional cut-constructible part is
given by
\be
M^{(2),D=4}=\frac{1}{16}\sum_{\sigma\in {\cal S}_1\cup {\cal S}_2\cup {\cal S}_3} 
 S^{(2),D=4}(\sigma)+{\cal O}(\epsilon)
\ee
where the coefficients $c_j$ in the identity permutation are given by
\be
\begin{array}{rlcrl}
c_1 =&
%%%%% begin : coeff1
s_{123} \bigl(s_{12} s_{45} s_{234}  + s_{23} s_{56} s_{345}
%%%%% end : coeff1
&~~~~~~&
c_2 =&
%%%%% begin : coeff2
2 s_{23} s_{12}^2 \\
%%%%% end : coeff2
&
%%%%% begin : coeff1
+s_{123} (s_{34} s_{61} - s_{234} s_{345})\bigr)
%%%%% end : coeff1
&~~&          &   \\
c_3 =&
%%%%% begin : coeff3
s_{123} (s_{345} s_{123} - s_{45} s_{12})
%%%%% end : coeff3
&&
c_4 =&
%%%%% begin : coeff4
s_{34} s_{123}^2 \\
%%%%% end : coeff4
c_5 =&
%%%%% begin : coeff5
s_{12} (s_{234} s_{123} - 2 s_{23} s_{56})
%%%%% end : coeff5
&&
c_6 =&
%%%%% begin : coeff6
- s_{61} s_{12} s_{123} \\
%%%%% end : coeff6
c_{7a} =&
%%%%% begin : coeff7a
 s_{123} (s_{234} s_{345}  - s_{34} s_{61})
%%%%% end : coeff7a
&&
c_{7b} =&
%%%%% begin : coeff7b
 - 4 s_{34} s_{61} s_{123} 
%%%%% end : coeff7b
\\
c_{7c} =&
%%%%% begin : coeff7c
 s_{123}( s_{234} s_{345} - s_{34} s_{61})
%%%%% end : coeff7c
&&
c_8 =&
%%%%% begin : coeff8
2 s_{12} (s_{345} s_{123} - s_{12} s_{45}) \\
%%%%% end : coeff8
c_9 =&
%%%%% begin : coeff9
s_{45} s_{56} s_{123}
%%%%% end : coeff9
&&
c_{10} =&
%%%%% begin : coeff10
s_{56} (2 s_{12} s_{45} - s_{123} s_{345}) \\
%%%%% end : coeff10
c_{11} =&
%%%%% begin : coeff11
s_{61} s_{56} s_{123}
%%%%% end : coeff11
&&
c_{12} =&
%%%%% begin : coeff12
s_{123} (s_{345} s_{123} - s_{12} s_{45}) \\
%%%%% end : coeff12
c_{13} =&
%%%%% begin : coeff13
- s_{123}^2 s_{61}
%%%%% end : coeff13
&&
c_{14} =&
%%%%% begin : coeff14
0 \\
%%%%% end : coeff14
c_{15} =&
%%%%% begin : coeff15
0
%%%%% end : coeff15
&&
c_{16} =&
%%%%% begin : coeff16
0\\
%%%%% end : coeff16
c_{17} =&
%%%%% begin : coeff14
-2 s_{123} s_{345} ( s_{234} s_{345} - s_{61} s_{34})
%%%%% end : coeff14
&&
c_{18} =&
%%%%% begin : coeff15
2 s_{12} ( s_{123} s_{234} s_{345} - s_{12} s_{45} s_{234})
%%%%% end : coeff15
\\
%%%%% end : coeff16
&
%%%%% begin : coeff14
+2 s_{345} ( s_{12}
  s_{45} s_{234} + s_{23} s_{56} s_{345})
%%%%% end : coeff14
&&
&
%%%%% begin : coeff15
-2 s_{12} ( s_{23}s_{56} s_{345} +s_{34} s_{61} s_{123})
%%%%% end : coeff15
\end{array}
\label{IntegralCoefficientsMHV}
\ee
%%%%%%%%%%%%%%%%%%%%%%%%%%%%%%%%%%%%%%%%%%

The $\mu$-integral contribution is
\be
M_6^{(2),\mu}=
\frac{1}{16} \sum_ {\sigma\in {\cal S}_1\cup {\cal S}_2\cup {\cal S}_3}
\biggl[
   \frac{1}{4} c_{17} I^{(17)}(\sigma)
+ \frac{1}{2} c_{18} I^{(18)}(\sigma)
\biggr]\,.
\ee
As mentioned previously, $\Int{17}$ starts at $\Ord(\eps)$
\cite{Bern:2008ap} and thus does not contribute through
$\Ord(\eps^0)$.

\subsection{A Comparison of the MHV and NMHV Amplitudes }

A direct inspection of the integrals in fig.~\ref{2loopints} and of
their coefficients in eq.~\eqref{IntegralCoefficientsNMHV}
reveals that the even part of the two-loop six-point NMHV amplitude is
a sum of pseudo-conformal integrals.
This is similar to the MHV amplitude, for which the four-dimensional
cut-constructible part has a similar property~\cite{Bern:2008ap}, as
may also be seen by directly inspecting the coefficients listed in
eq.~\eqref{IntegralCoefficientsMHV}.

This is perhaps not completely surprising in light of the argument
presented in section~\ref{higher_loop_structure} that all
four-dimensional cuts can be reproduced by cuts of pseudo-conformal
integrals.  This structure does not guarantee, however, that the even
part of the amplitude is dual conformal invariant, even after
infrared divergences are removed appropriately. We return to this
point in the next section.

The structure of the NMHV amplitude is quite similar to that of the
MHV amplitude, with only subtle differences in the values of the
coefficients. Two of the integrals
that did not contribute to the MHV amplitude --- $\Int{14}(\eps)$
and $\Int{16}(\eps)$ --- enter in the NMHV amplitude with
nonvanishing coefficient; similarly, a topology that exists in both
amplitudes --- $\Int{1}(\eps)$ --- appears in the NMHV amplitude with
an additional pseudo-conformal numerator. Moreover, an integral that
contributes to the MHV amplitude --- $\Int{8}(\eps)$ --- disappears
from the NMHV one.
We note also that a perfectly valid integral --- $\Int{15}(\eps)$ --- appears
in neither the MHV nor NMHV amplitudes. It would be
interesting to understand the significance of this observation.

The properties of the MHV and NMHV amplitudes differ from those observed 
in the four-point amplitudes through five loops:
\begin{itemize}
\item All pseudo-conformal integrals
appear with relative weights of $\pm 1$ or
$0$~\cite{BRY,ABDK,BCDKS,FiveLoop}.

\item
An integral appears with coefficient zero if 
and only if the integral is unregulated
after taking its external legs off shell
and taking $\epsilon \rightarrow 0$~\cite{DKS}.

\item
It has been proposed that the signs $\pm 1$ of the
contributing integrals can be understood by the requirement
of cancelling unphysical singularities~\cite{CachazoSkinner}.
\end{itemize}
It would undoubtedly be interesting to understand the generalization
of these features to higher-multiplicity amplitudes.

Although the amplitudes have very similar structures, the ratio of 
the NMHV six-point superamplitude to the MHV one does not appear to exhibit a
transparent organization. This is due to the
different structures of the permutation sums that contribute to the
independent factors in the amplitudes. Indeed, each function
$W_i^{(1),D=4}$ and $W_i^{(2),D=4}$ contains only a sum over the permutations in
the set ${\cal S}_i$ while the functions $M^{(1),D=4}$ and
$M^{(2),D=4}$ contain sums over all twelve permutations ${\cal S}_1\cup {\cal
  S}_2\cup {\cal S}_3$.

%%%%%%%%%%%%%%%%%%%%%%%%%%%%%%%%%%%%%%%%%

\section{Dual Conformal Invariance \label{results} }

The explicit calculation in section~\ref{TheCalculation} of the
two-loop NMHV six-point superamplitude shows that it indeed has the
structure anticipated in section~\ref{higher_loop_structure}.  We
obtained explicit integral representations for the scalar functions
$W_i^{(2)}$, summarized in the previous section.

As it is the case with all massless theories in four dimensions, the
amplitude is infrared divergent; to examine the dual conformal
properties of the amplitude it is necessary to isolate these
divergences. Based on the universality of infrared singularities and
their exponentiation DHKS proposed~\cite{Drummond:2008vq} that these
divergences be removed by simply dividing by the MHV amplitude. This
ratio (\ref{eq:all-loop-nmhv}) was conjectured to be dual conformally
invariant. An alternative method was described in
section~\ref{higher_loop_structure}, see eq.~\eqref{eq:Witeration}.
%%%%%%%%%%%%%%
% Clearly, the remainder-like functions $R_{6;i}^{(2)}$ and the DHKS functions 
% $\Coeff_{i,i+3i+5}$ are related.
%%%%%%%%%%%%%%

Since the functions $W_i^{(l)}$ have a natural decomposition into four-dimensional 
and $D$-dimensional cut-constructible contributions, the functions $\Coeff_i^{(l)}$ introduced 
in equation (\ref{new_ratios}) inherit a similar decomposition
\be
W_i^{(l)}=W_i^{(l),D=4}+W_i^{(l),\mu}
~~~
\longrightarrow
~~~
\Coeff_i^{(l)}=\Coeff_i^{(l),D=4}+\Coeff_i^{(l),\mu}~~~~l=1,2~~.
\ee
At one loop, the $\mu$-integral contribution $\Coeff_i^{(1),\mu}$
vanishes in the limit $\eps\rightarrow 0$.  Because of infrared divergences, they
nevertheless give rise to nontrivial contributions at two loops in the
ratio with the MHV amplitude.
The functions 
$\Coeff_i^{(l)}$ contain terms that vanish as $\epsilon\rightarrow 0$:
\be
\Coeff_i^{(1)}(\epsilon)&=&W_i^{(1)}(\epsilon)-M^{(1)}(\epsilon)
\cr
\Coeff_i^{(2)}(\epsilon)&=&W_i^{(2)}(\epsilon)-M^{(2)}(\epsilon)
                                 -M^{(1)}(\epsilon)(W_i^{(1)}(\epsilon)-M^{(1)}(\epsilon))~~.
\label{Vfcts}
\ee
As explained in sect.~\ref{SixPointNMHVSuperamplitudeSection},
in the limit $\eps\rightarrow 0$, $\Coeff_i^{(1)}(\eps)$ 
reduces to dual conformal
invariant functions very closely related to 
the $V^{(i)}$ defined in ref.~\cite{Drummond:2008vq}. 
The higher-order terms in $\eps$ were not considered in 
ref.~\cite{Drummond:2008vq}; we take them as 
determined by the amplitude through the
relation~\eqref{new_ratios} between $W_i^{(l)}$ and $\Coeff_i^{(l)}$.

The infrared-divergent terms in both $W_i^{(1)}(\epsilon)$ and
$M^{(1)}(\epsilon)$ have the usual form
\begin{equation}
  \text{Div}_6^{(1)} = - \frac 1 {2 \epsilon^2} \sum_{j=1}^6 (-s_{j,j+1})^{-\epsilon}~~.
\label{div1loop6}
\end{equation}  
Thus, $\Coeff_i^{(1)}(\epsilon)$ is manifestly finite; moreover, the
only divergent contribution arising from the last term in
$\Coeff_i^{(2)}(\epsilon)$ is due to the overall factor of
$M^{(1)}(\epsilon)$.

\subsection{Terms Requiring $D$-dimensional Cuts}

At one loop, the $\mu$-integrals, requiring consideration of
$D$-dimensional cuts, yield only terms of $\Ord(\eps)$ in both
$W_i^{(1)}(\epsilon)$ and $M^{(1)}(\epsilon)$.  This allows us to
isolate the $\mu$-integral contribution in eq.~(\ref{Vfcts}):
\begin{equation}
  \Coeff_i^{(2),\mu} = 
W_i^{(2),\mu} - W_i^{(1),\mu} \text{Div}_6^{(1)}
- M^{(2),\mu} +  M^{(1),\mu} \text{Div}_6^{(1)} + \mathcal{O}(\epsilon)\,,
\end{equation} 
where we used the universality of the one-loop infrared divergences
(\ref{div1loop6}) and kept only the terms that have nontrivial
divergent and finite parts. For example, we dropped the terms in
eq.~\eqref{Vfcts} coming from the finite part of the overall factor
$M^{(1)}(\eps)$ in the last term.

The last two terms in the equation above contain information already
available in the MHV amplitude. Indeed, this exact combination appears
in the iteration of the $\mu$-integrals for this amplitude~\cite{Bern:2008ap}:
\be
M^{(2),\mu}=M^{(1),\mu} \text{Div}_6^{(1)}\,.
\ee
Thus, $\Coeff_i^{(2),\mu}$ is given by,
\begin{equation}
\label{VmuF}
  \Coeff_i^{(2),\mu} = W_i^{(2),\mu} - W_i^{(1),\mu} \text{Div}_6^{(1)}\,,
\end{equation}
with $W_i^{(2),\mu}$ given by eq.~\eqref{mu_ints_W}.

This expression for $W_i^{(2),\mu}$ may be further simplified by making
use of the special properties of the hexabox integral discussed in
section~IV.A of ref.~\cite{Bern:2008ap}, in particular eq.~(4.6):
\be
\Int{18}[\mu^2]=-\frac{1}{\epsilon^2}(-s_{12})^{-1-\epsilon}I^{\rm hex}[\mu^2]~~.
\ee
Thus, $W_i^{(2),\mu}$ can be expressed exactly in terms of one-loop
integrals, albeit in six dimensions. Moreover, using the fact that the
one-loop hexagon integral is invariant under cyclic permutations of
external legs, $W_i^{(2),\mu}$ can be expressed in terms of the
massless six-dimensional hexagon integral with a coefficient given by
the universal divergent part of one-loop amplitudes:
\begin{equation}
\label{WmuF}
  W_i^{(2),\mu} = -\frac{1}{12} \text{Div}_6^{(1)} I^{\rm hex}[\mu^2] \sum_{\sigma\in {\cal S}_i} 
  s_{123} (2 s_{34} s_{61} - s_{234} s_{345})   + \mathcal{O}(\epsilon)~~.
\end{equation}  
The four terms in each sum are in fact equal.

The $\mu$-integral contribution to the one-loop six-point NMHV
amplitude is not yet available in the literature.  Information on its
structure may be obtained by analyzing a two-particle cut of the
$\mu$-integral contribution to the two-loop NMHV superamplitude. Dixon
and Schabinger~\cite{SD} have evaluated such a cut directly; quite
surprisingly, they find that it can be organized in terms of the same
$R$ invariants as the four-dimensional cut-constructible terms.  The
$\mu$-integrals' contribution to the even part of the one-loop
six-point NMHV amplitude is, 
\be
\label{1loop_mu}
W_i^{(1),\mu} = -\frac{1}{12} I^{\rm hex}[\mu^2] \sum_{\sigma\in {\cal S}_i} 
  s_{123} (2 s_{34} s_{61} - s_{234} s_{345}) ~~.
\ee
Combining this with eqs.~\eqref{WmuF} and~\eqref{VmuF} immediately shows
that
\begin{equation}
  \lim_{\epsilon \to 0} \Coeff_i^{(2), \mu} = 0~~.
  \label{zeroVmu}
\end{equation}  
In other words, the complete $\mu$-integral contribution to the
six-point two-loop NMHV amplitude is completely accounted for by
extracting an overall factor of the MHV superamplitude.

Through similar manipulations it is possible to show that the
remainder-like functions $R_{6;i}^{(2)}$ introduced in
eq.~\eqref{eq:Witeration} do not receive any $\mu$-integral
contributions. Indeed, directly expanding eq.~\eqref{eq:Witeration} to
$\Ord(a^2)$ we find that,
\be
R_{6;i}^{(2)}=W_i^{(2)}(\epsilon)-\left[\frac{1}{2}\left(W_i^{(1)}(\epsilon)\right)^2
+f_2(\epsilon)W_i^{(1)}(2\epsilon)\right]~~.
\ee
Identifying the $\mu$-integral contributions to each of the terms on
the right hand side and using the universality of infrared divergences
implies that 
\be
R_{6;i}^{(2),\mu}=W_i^{(2),\mu}(\epsilon)-\text{Div}_6^{(1)}W_i^{(1)}(\epsilon)
=\Coeff_i^{(2),\mu} +{\cal O}(\epsilon)~~.  \ee It therefore follows
from equations~(\ref{zeroVmu}) that $R_{6;i}^{(2),\mu}$ does not
receive any finite $\mu$-integral contributions.

The same, however, cannot be said about the $\mu$-integral
contribution to the odd part of the amplitude. Indeed, repeating
the steps that lead to equations~(\ref{Vfcts}) we find that
the coefficients of the parity-odd quantities
$(R_{i+3,i,i+2}-R_{i,i+3,i+5})$ are,
\be 
\Coefftilde_i^{(2)}={\widetilde W}_i^{(2)} - \text{Div}_6^{(1)}\,{\widetilde
  W}_i^{(1)}~~.  \ee 
While the $\mu$-integral contributions to
${\widetilde W}_i^{(2)}$ are given in terms of the hexabox integral or,
equivalently in terms of the six-dimensional hexagon integral, their
contributions to ${\widetilde W}_i^{(1)}$ are given in terms of a restricted
set of the one-mass pentagon integrals~\cite{SD}.  This suggests that, for the
odd part of the superamplitude, the $\mu$ integrals cannot be cleanly
separated from the four-dimensional cut-constructible terms.

\subsection{Numerical Evaluation of the Amplitude}

In order to further analyze the properties of the two-loop six-point NMHV
amplitude, we turn to a numerical evaluation of the two-loop
integrals. Thanks to the results described in the previous section, we
may focus on the four-dimensional cut-constructible part of the
amplitude.
The task of evaluating the integrals is simplified substantially by
the fact that all of them have already been evaluated at several
distinct kinematic points in~\cite{Bern:2008ap}. 
We have evaluated additional kinematic
points using the package {\tt MB}~\cite{MB} and the same
Mellin--Barnes parametrization of integrals that was used in the
calculation of the MHV amplitude.
Apart from testing the symmetry properties of the amplitude, this
calculation also verifies the expected universality of two-loop
infrared divergences.  Viewed differently, a successful test of the
universality of the infrared divergences is a strong indication of the
completeness of the cut construction described in previous sections.

We choose Euclidean kinematics for all configurations of external
momenta. As in the calculation of the MHV amplitude, the
symmetries of the momentum configuration,
\be K^{(0)}:
~~~~s_{i,i+1}=-1\,,\;s_{i,i+1,i+2}=-2\,,
\ee 
make it particularly
useful, as all cyclic permutations or external legs yield the same
value for all integrals. This implies that all functions
$W^{(2),D=4}_{i}$ are equal for all $i=1,2,3$.  Using the values of
the integrals collected in the Appendix~B of~\cite{Bern:2008ap} we
find \be W^{D=4}_{i}(K^{(0)})\!&=&\!  1 + a( - \frac{3}{\epsilon^2}
+5.27682+ 8.73314 \epsilon + 8.11147 \epsilon^2) \nonumber\\[3pt]
&&~\, + a^2 \Big(\frac{9}{2\epsilon^4}-\frac{14.5967}{\epsilon^2}
+\frac{25.3014\pm 0.0043}{\epsilon}-21.064\pm 0.002 \Big)+{\cal
  O}(a^3).~~~~~~~~
\label{WiK0}
\ee
Where they are not explicitly included, the errors do not affect the
last quoted digit. We have used the error estimated reported by {\tt
  CUBA}~\cite{CUBA}. In general we found the errors to be reliable,
giving an accurate measure of the number of trustworthy digits. In
some contributions, however,
 we found them to be underestimated, invariably in the
presence of small integrals with a fast-varying integrand. In such
cases, when {\tt CUBA} reports a large $\chi^2$, we take the average
value of the integrals to be the central value and quote the variation of the
integral under changes of sampling points as the error estimate. 
The issue presumably involves integration regions missed because of
special properties of the integrand.

As discussed previously, the construction of the functions
$\Coeff_i^{(2)}$ requires keeping higher orders in the 
small-$\eps$ expansion of the one-loop amplitude. For the MHV amplitude,
we use the expression in terms of the iterated one-loop amplitude and
the remainder function $R_6^{(2)}$.
For the point $K^{(0)}$ we find
\be
\Coeff_i(a,\epsilon,K^{(0)})
%&=&\frac{W_{i}}{M_6}=\frac{W^{D=4}_{i}}{M_6^{D=4}}
%\\
&=&
1+ a (0.783676 + 1.10087 \epsilon + 0.07507 \epsilon^2)
\\
&&~~
 +a^2 \Big(- \frac{0.0036\pm0.0043}{\epsilon} -(2.412\pm 0.002)-R_6^{(2)}(K^{(0)}) \Big)
 +{\cal O}(a^3).
\nn
\ee
We note that the residue of the simple pole in $\epsilon$ vanishes
within errors, as it should.  We have confirmed this property for all
the other kinematic points\footnote{We have also verified
  analytically the cancellation of infrared singular terms through
  $\Ord(\epsilon^{-2})$. The complete cancellation of
  infrared-singular terms was shown analytically by G.~Korchemsky
  (private communication).}.

From equation~(\ref{WiK0}) we can also find the value of the
remainder-like functions $R_{6;i}^{(2)}$ introduced in equation
(\ref{eq:Witeration}) at the point $K^{(0)}$: \be
R_{6;i}^{(2)}=-1.430\pm 0.002~~.  \ee Similarly to the error quoted
for $\Coeff_i(a,\epsilon,K^{(0)})$, the error of $R_{6;i}^{(2)}$ is
completely inherited from that of $W_i^{(2)}$.

We have evaluated the amplitude at another kinematic point (denoted by $K^{(1)}$ in 
\cite{Bern:2008ap}) related to $K^{(0)}$ by dual conformal transformations as well as two 
other points related to each other but unrelated  to $K^{(0)}$:
\be
%
%  kin2**
%
K^{(1)} &:&  s_{12} = -0.723 6200, \hs  s_{23} = -0.921 3500, \hs
 s_{34} = -0.272 3200, \hs  s_{45} = -0.358 2300, \hs \nn \\
&&
 s_{56} = -0.423 5500, \hs  s_{61} = -0.321 8573,  \hs
 s_{123} = -2.148 6192,  \hs  s_{234} = -0.726 4904, \hs \nn \\
&&
 s_{345} =  -0.482 5841,  \nn \\
 %
% kin4** 
%
K^{(3)} &:& s_{i,i+1} = -1,  \hs  s_{123} = -1/2,  \hs s_{234} = -5/8, \hs 
             s_{345} = -17/14, \nn \\
K^{(6)} &:& s_{12} = -2, \hs  s_{23} = -4, \hs
 s_{34} = -2 , \hs  s_{45} = -14/17, \hs s_{56} = -4/5, \hs  s_{61} = -56/85,  \hs
 \nn \\
&&
 s_{i,i+1,i+2} = -1, \hs
\ee
In listing the kinematic points we attempted to preserve the notation
for the points used in ref.~\cite{Bern:2008ap}. We have collected our
results for the values of $\Coeff_i^{(2)}$ and $R_{6;i}^{(2)}$ in
tables~\ref{comparisonV} and~\ref{comparisonR}.  The three dual
conformal ratios
\be
(u_1,u_2,u_3)=
\Big(\frac{s_{12}s_{45}}{s_{123}s_{345}},\frac{s_{23}s_{56}}{s_{234}s_{456}},
\frac{s_{34}s_{61}}{s_{345}s_{561}}\Big)
\ee
for the kinematic points are listed in the second column of these tables.

\begin{table}[h]
\caption{\label{comparisonV}Comparison of conformally-related kinematic points. $\Coeff_i^{(2)}$
are the finite parts of the ratios $\Coeff_i=W_i/M_6$ at two-loops. $R_6^{(2)}$ is the two-loop 
remainder function of the six-point MHV amplitude.}

\begin{center}
\begin{tabular}{||c|c|c|c|c||}
\hline
\hline
kinematic pt.            &$(u_1,u_2,u_3)$      & $\Coeff_1^{(2)}+R_6^{(2)}$ & $\Coeff_2^{(2)}+R_6^{(2)}$ &$\Coeff_3^{(2)}+R_6^{(2)}$
\\[1pt]
\hline
\hline
$K^{(0)}$   &$(\frac{1}{4},\frac{1}{4},\frac{1}{4})$& 
$-2.413\pm 0.002$& $-2.413\pm 0.002$& $-2.413\pm 0.002$\\[1pt]
\hline
$K^{(1)}$   &$(\frac{1}{4},\frac{1}{4},\frac{1}{4})$& 
$-2.359\pm 0.048$&$-2.375\pm 0.025$&$- 2.380\pm 0.033$ \\[1pt]
\hline
\hline

$K^{(3)}$   &$(\frac{28}{17}, \frac{16}{5}, \frac{112}{85})$& 
$14.426\pm 0.003$& $12.614\pm 0.004$& $11.697\pm 0.009 $\\[1pt]
\hline
$K^{(6)}$   &$(\frac{28}{17}, \frac{16}{5}, \frac{112}{85})$& 
$ 14.439\pm 0.078$&$ 12.614\pm 0.035 $&$ 11.727\pm 0.145$ \\[1pt]
\hline\hline
\end{tabular}
\end{center}
\end{table}

\begin{table}[h]
\caption{\label{comparisonR} Comparison of the remainder-like functions $R_{6;i}^{(2)}$ at 
conformally-related kinematic points. }

\begin{center}

\begin{tabular}{||c|c|c|c|c||}
\hline
\hline
kinematic pt.            &$(u_1,u_2,u_3)$      & $R_{6;1}^{(2)}$ & $R_{6;2}^{(2)}$ &$R_{6;3}^{(2)}$
\\[1pt]
\hline
\hline
$K^{(0)}$   &$(\frac{1}{4},\frac{1}{4},\frac{1}{4})$& 
$-1.431\pm 0.002$& $-1.431\pm 0.002$& $-1.431\pm 0.002$\\[1pt]
\hline
$K^{(1)}$   &$(\frac{1}{4},\frac{1}{4},\frac{1}{4})$& 
$-1.377\pm 0.048$&$-1.393\pm 0.025$&$- 1.397\pm 0.033$ \\[1pt]
\hline
\hline
$K^{(3)}$   &$(\frac{28}{17}, \frac{16}{5}, \frac{112}{85})$& 
$5.413\pm 0.003$& $4.749\pm 0.004$& $4.602\pm 0.009 $\\[1pt]
\hline
$K^{(6)}$   &$(\frac{28}{17}, \frac{16}{5}, \frac{112}{85})$& 
$ 5.427\pm 0.078$&$ 4.749\pm 0.035 $&$ 4.633\pm 0.145$ \\[1pt]
\hline
\end{tabular}
\end{center}
\end{table}

As mentioned previously, we left the MHV remainder function $R_6^{(2)}$ 
unevaluated. Its dual conformal invariance 
\cite{Bern:2008ap} 
\be
R_6^{(2)}(K^{(0)})=R_6^{(2)}(K^{(1)})
~~~~~~~~~
R_6^{(2)}(K^{(3)})=R_6^{(2)}(K^{(6)})
\ee
implies that the equality within errors of the relevant entries of
table~\ref{comparisonV} extends to an equality of the functions
$\Coeff_i^{(2)}$. Alternatively, we could have evaluated the remainder
function from the analytic expression found in 
ref.~\cite{DelDuca:2009au}, the integral representation in
ref.~\cite{Zhang}, or the
simplified form in ref.~\cite{Goncharov:2010jf}.
The results we obtain thus suggest that $\Coeff_i^{(2)}$ and
$R_{6;i}^{(2)}$ are functions solely of the conformal cross-ratios, that is,
that they are indeed invariant under dual conformal transformations.

\section{Summary, conclusions and some open questions \label{summary} }

The maximally supersymmetric gauge theory in four dimensions is an
ideal testing ground for probing the properties of gauge theories at
both weak and strong coupling. The large degree of symmetry makes
perturbative calculations tractable to relatively high orders while
its string-theory dual provides powerful tools for understanding its
strong-coupling behavior.  Its hidden symmetries yield additional
constraints that go beyond their initial connection to the
integrability of the dilatation operator of the theory.

In this paper we have computed the parity-even part of the two-loop
six-point NMHV amplitude 
using generalized unitarity in
superspace.  We showed that
 the result is invariant under dual
conformal transformations,
after removal of universal
infrared divergences (including terms arising from $\Ord(\eps)$ 
contributions at one loop, computed by Dixon and Schabinger~\cite{SD}).
 The dual conformal invariant content
may be organized in several different ways. The exponentiation of both
the infrared divergences and of the collinear splitting amplitudes
suggest the introduction of certain remainder-like functions which,
similarly to the remainder function for MHV amplitudes, are
functions only of the conformal cross ratios.  We have shown that, to
all orders in perturbation theory, it should be possible to
reconstruct the remainder-like functions by evaluating certain
triple-collinear splitting amplitudes.

Several interesting issues related to the calculation described here,
and to the structure of the perturbative expansion of the theory and
its strong coupling expansion remain to be clarified.

Through the AdS/CFT correspondence, Alday and Maldacena~\cite{AM1}
argued that, to leading order in their strong-coupling expansion, all
planar scattering amplitudes with fixed number of external legs are
essentially identical up to perhaps a rational function of momenta and
polarization vectors. Our calculation and arguments show that the
weak-coupling structure of the six point amplitude involves at least
six different spin factors dressed with scalar and pseudo-scalar
functions to all orders in the weak-coupling
expansion. Reconciling this structure with the AdS/CFT considerations
remains an important open problem, which appears to require that, in
the strong-coupling limit, the remainder-like functions introduced in
section~\ref{higher_loop_structure} have a very simple relation to the
MHV remainder function.

Inspired by strong-coupling considerations, several groups showed
that MHV
amplitudes have a close relation to certain light-like polygonal Wilson loops
order-by-order in weak-coupling perturbation 
theory, first in explicit calculations~\cite{DKS,BHT,DHKS2loops,Bern:2008ap},
and very recently, cast in a more general setting~\cite{WilsonCorrelationAmplitudeLink}.
A similar
relation for non-MHV amplitudes remains an open question; our
numerical results provide check points for future calculations in this
direction. The Wilson-loop formulation of the six-point MHV amplitude
led to the analytic evaluation~\cite{DelDuca:2009au,Zhang,Goncharov:2010jf} 
of the remainder
function at two loops. It seems likely that a Wilson-loop formulation
of NMHV amplitudes will allow a similar evaluation of the
remainder-like functions characterizing this amplitude.

A direct comparison of the integrands of the six-point MHV and NMHV
amplitudes at two loops reveals that certain integrals appear in one
but not the other, while the contributing integrals enter with
numerical coefficients that do not conform with the effective rules
inferred from four-point amplitudes. Moreover, one perfectly valid
pseudo-conformal integral does not appear in either one of these
amplitudes. It would be interesting to develop a better understanding
of this pattern of numerical coefficients. Evaluation of higher-point
NMHV amplitudes at two loops may help in this direction.

In our calculation dual conformal invariance is obscured by the
dimensional regulator; removal of infrared divergences is a
crucial step in studying the dual conformal properties of scattering
amplitudes. Using four and five-point amplitudes as testing ground, it
was shown~\cite{AHPS, massregulator} that regulating the infrared
divergences by a particular symmetry breaking of the gauge
group makes dual conformal invariance more transparent. It would be
interesting to repeat the calculation described in this paper as well
as that of the two-loop six-point MHV amplitude in this
framework. Apart from a better understanding of dual conformal
invariance, such an endeavor would clarify the interpretation of
$\mu$-integrals as well as that of the parity-odd component of
amplitudes. While we did not compute the parity-odd
part of the six-point NMHV amplitude explicitly, its $\mu$-integral
contributions make it quite different from the parity-odd terms in
MHV amplitudes with up to six external legs.
     
A very interesting question relates to symmetries of scattering
amplitudes. Up to anomalies introduced by the regulator, it is
expected that dual conformal transformations leave scattering
amplitudes invariant to all orders in perturbation theory. Again apart
from anomalies due to the presence of a regulator, ordinary conformal
invariance exhibits additional anomalies of a holomorphic type
\cite{CSWholomorphic}, related to singular momentum configurations,
already at tree level. An appropriate definition of a generating
function for superamplitudes with variable number of external legs
\cite{BeisertConformal} allows this anomaly to be circumvented, and to
be realized on scattering amplitudes through the one-loop level.
Korchemsky and Sokatchev have recently given~\cite{Korchemsky:2010ut}
a general construction of conformal and dual conformal invariants (and
hence of Yangian invariants). Other approaches to the construction
of Yangian invariants were discussed 
by Mason and Skinner~\cite{Mason:2009qx} and
by Drummond and Ferro~\cite{Drummond:2010uq}.  
These invariants have expressions that
were conjectured~\cite{ArkaniHamed:2009dn} by Arkani-Hamed {\it et
  al.\/} to represent the leading singularities of scattering
amplitudes to all orders in perturbation theory. All-order leading singularities 
have been derived directly by
Bullimore, Mason, and Skinner~\cite{Bullimore:2009cb}.
While the form of
subleading singularities is not yet clear, this structure suggests
that it may be possible to realize both symmetries (and their closure)
at higher loops.  The extended algebra does not uniquely determine the
S-matrix of \NeqFour{} super-Yang-Mills theory,
however~\cite{Korchemsky:2010ut}.  Unraveling its constraints
on scattering amplitudes should prove fruitful.

The study of non-MHV amplitudes at higher loops is only in its early
stages. Our explicit calculation provides an example of such an
amplitude. The structure of these amplitudes is substantially richer
than that of MHV amplitudes. It seems likely that new and exciting
properties as well as new calculational techniques are waiting to be
discovered.

\section*{Acknowledgments}

We are grateful to Zvi Bern, Lance Dixon, Gregory Korchemsky, Robert
Schabinger, Emery Sokatchev, and Arkady Tseytlin for useful
discussions. We are especially grateful to Lance Dixon and Robert
Schabinger for sharing their results on the $\mu$-integral
contribution to the one-loop six-point NMHV amplitude prior to
publication.  We thank Marcus Spradlin and Zvi Bern for help with
integral numerics.
This work was supported in
part by  the European Research Council under Advanced 
Investigator Grant ERC--AdG--228301 (D.~A.~K.), 
the US Department of Energy under 
contracts DE--FG02--201390ER40577 (OJI) (R.~R.)
and DE--FG02--91ER40688 (C.~V.), the US National Science Foundation under
PHY--0608114 and PHY--0855356 (R.~R.) and PHY--0643150 (C.~V.)
and the A.~P.~Sloan Foundation (R.~R.).
We also thank Academic Technology Services at UCLA for computer support.
Several of the figures were
generated using Jaxodraw~\cite{Jaxo} (based on Axodraw~\cite{Axo}).

%%%%%%%%%%%%%%%%%%%%

\end{document}